\documentclass[reprint,amsmath,amssymb,aps,twocolumm,superscriptaddress,longbibliography]{revtex4-1}
\usepackage[utf8]{inputenc}
\usepackage[T1]{fontenc}
\usepackage{graphicx}
\usepackage{dcolumn}
\usepackage{bm}
\usepackage[colorlinks,linkcolor=blue,citecolor=blue,urlcolor=blue,breaklinks=treu]{hyperref}

\begin{document}

\title{Frequency Multiplexed Optical Entangled Source based on the Pockels Effect}
\author{Alfredo Rueda}
\affiliation{Scantinel Photonics GmbH, Hirschstrasse 4, 89073 Ulm Germany}
\affiliation{Institute of Science and Technology Austria, am Campus 1, 3600 Klosterneuburg, Austria}
\date{\today}

\begin{abstract}
In the recent years important experimental advances in resonant electro-optic modulators as high efficient sources for coherent frequency combs and as devices for quantum information transfer  have been realized, where strong optical and microwave mode coupling were achieved. These features suggest electro-optic based devices as candidates for entangled optical frequency comb sources. In the present I study the generation of entangled optical frequency combs in mm-sized resonant electro-optic modulators. These devices profit from the experimentally proven advantages such as nearly constant optical free spectral ranges over several gigahertz, high optical and microwave quality factors. The generation of frequency multiplexed quantum channels with spectral bandwidth in the range of a MHz for conservative parameter values paves the way towards novel uses in long distant hybrid quantum networks, quantum key distribution, enhanced optical metrology and quantum computing.
\end{abstract}

\pacs{Valid PACS appear here}\maketitle

\section{Introduction}
In the last years remarkable breakthroughs towards a global quantum computation network have been made~\cite{Geoff,Wendin_2017}. Quantum processors based on superconducting microwave circuits are now the leading platform for quantum computers, which has already achieved supremacy over their classical counterparts~\cite{google_1}. The next level in the quantum technologies is interconnection, which can only be given by a quantum network, that can harness all the advantages of quantum correlations. A full microwave quantum network is not feasible, since the whole system must be cooled down to mK regime to protect the weak microwave photons from environmental thermal excitations \cite{Kupiers2018,Chou2018}. On the other hand, at optical frequencies middle and long distance quantum channels and networks are feasible and their capabilities have been tested with promising results \cite{Liao2018,luo2019}. 

At this point the need of a  functional hybrid quantum network rises as key ingredient for the success of global quantum computing~\cite{Kimble2008,Maring17}, making possible the integration of advanced microwave quantum state processing~\cite{Hofheinz2009, Eichler2011a, Vlastakis2013} with the well developed optical discrete variable (DV) and continuous variable (CV) quantum information protocols~\cite{Ulricht,weedbrook2012} such as quantum state teleportation~\cite{Furusawa706,Lee330} and dense coding \cite{mattle96,jing2003}.

The current paradigm to integrate a microwave quantum state into well stablished  optical quantum networks~\cite{Guccioneeaba4508,spiropulu2020} is through coherent frequency up-conversion (CFC). The two most successful systems are currently based on the electro-optomechanical (EOM) and electrooptic (EO) approach \cite{Lambert2019}. The former offers high conversion efficiency $\sim50\%$ but it suffers of modest conversion bandwidth $\sim10$ kHz and high excess of added noise \cite{higgin} for faithful transduction. The latter offers a broadband conversion $\sim10$ MHz and low noise~\cite{willy2020}, but a conversion efficiency of 2\% has been so far demonstrated \cite{FanZouChengEtAl2018}. In this way, information and purity of the state of the initial states will degrade due to the imperfections of the converters~\cite{rueda2019electrooptic}. Therefore,  the need of a hybrid source for quantum network with microwave and optical channels would allow direct interconnection between the different systems, avoiding the need of microwave frequency up-conversion to access to the common network. EOM and EO-systems are theoretically capable of generation of electro-optic entanglement, which generates a hybrid quantum channel itself. EOM entanglement has been theoretically studied~\cite{Wang2013, Tian2013, Zhong2019} but its experimental realization remains challenging due the high optically induced the added noise.

The suitability of EO modulators  to generate a hybrid  single quantum channel has been studied recently in Ref.~\cite{rueda2019electrooptic}. Furthermore, in an mm-sized electro-optic system, there is the possibility of generating $N$-hybrid single quantum channels interconnected through a common microwave mode.
In the present work, I study the entanglement creation of one optical mode to a microwave mode and a second optical field. Further, I extent this study to multiple optical modes and I propose a cavity electro-optic modulator based on a mm-sized multiresonant whispering gallery mode (WGM),  whose free spectral range matches the microwave resonance frequency as a source for a electro-optic quantum network. Furthermore, I present the theory to analytically predict and quantify entanglement properties under realistic conditions like losses, asymmetric modes, internal and external thermal noise. I show and discuss with experimentally proven parameters to deterministically generate MHz bandwidth CV entanglement between a microwave mode and frequency multiplexed optical modes  in cryogenic environments via spontaneous parametric down conversion (SPDC) and coherent frequency upconversion (CFC) in a single device. I also present the performance of these entangled modes as quantum channels by estimating fidelity values for teleportation-based communication for a set of typical quantum states and its enhanced capacity over classical channels. Finally, the presented analytical results applies to systems such as magneto-optics\cite{hisatomi}, $\Lambda$-system~\cite{xavi2}, electro-mechanics~\cite{shabinature}, among others.


\section{General Electro-optic two optical mode entanglement}
\begin{figure}[t]
	\centering
		\includegraphics[width=0.45\textwidth]{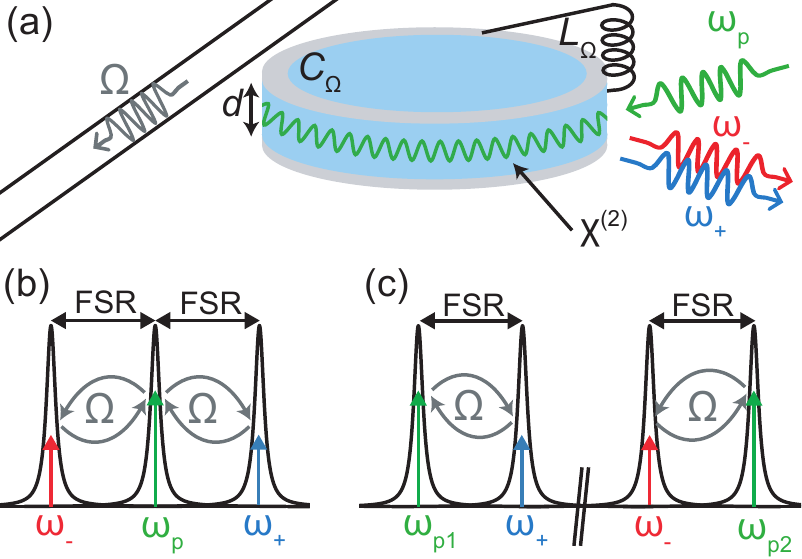}
	\caption{Schematic representation of the cavity electro-optic modulator. \textbf{(a)} A $\chi^{(2)}$-nonlinear optical resonator is confined between two metallic electrodes forming the capacitance $C_\Omega$ of a microwave resonator with resonance frequency $\Omega=1/\sqrt{L_\Omega C_\Omega}$. A coherent optical field at $\omega_p$ creates two optical and one microwave correlated field. Efficient  interaction requires matching $\Omega$ with the free spectral range of the optical mode.  \textbf{(b)} Driving scheme for symmetric FSR systems.  An optical mode is coherently pumped on resonance with  $\omega_p$ and the generation of two optical and one microwave fields is given through spontaneous parametric down conversion and coherently frequency upconversion. The resonantly created signals' frequencies corresponds the adjacent optical modes $\omega_{-}$ (Stokes),  $\omega_{+}$ (anti-Stokes) and the microwave mode $\Omega$. \textbf{(c)} Separated pumping scheme. One of the isolated two-optical-mode systems in the resonator is pumped with $\omega_{p1}$ (red-detuned) and the other with $\omega_{p2}$ (blue-detuned), achieving the SPDC and CFC through the same microwave mode $\Omega$. }
	\label{setupsim}
\end{figure}
Interaction between several electromagnetic waves occurs in media with nonlinear electric polarizability. The Pockel's effect describes the interaction between an optical wave $\omega_p$ and a microwave $\Omega$ in media with quadratic response to the electric field. This interaction generates two optical signals, referred to as the anti-Stokes and Stoke modes with frequencies $\omega_\pm=\omega_p\pm\Omega$, respectively \cite{amnon1989quantum,RuedaSanchez2018}. This process is described by the interaction Hamiltonian\cite{ilchenko_whispering-gallery-mode_2003, tsang_cavity_2011,Rueda:16}:
\begin{equation}
\hat{H}_\text{int}=\hbar g_+ \hat{a}_+^\dagger\hat{a}_\Omega \hat{a}_p+\hbar g_- \hat{a}^\dagger_-\hat{a}_\Omega^\dagger \hat{a}_p+ H.c, \label{hamscat}
\end{equation}
with $a_k$ ($a_k^\dagger$) stand for the annihilation (creation) operator for the pump mode ($k=p$), the anti-Stoke mode ($k=+$), Stokes mode ($k=-$) and the microwave mode ($k=\Omega$), respectively. The first term describes the creation of a photon with frequency $\omega_+$ and the annihilation of two photons with frequencies $\omega_p$ and $\Omega$. The second term accounts for the decay of a photon in mode $\omega_p$ into two photons in modes $\omega_-$ and $\Omega$, respectively. 
The strength of the nonlinear interaction, given by the single photon coupling rate $g_\pm$, is determined by the spatial overlap between the participating waves and the medium effective electro-optic coefficient $r$~\cite{Rueda:16}. I limit the analysis to the case of a strong optical drive at the frequency $\omega_p$ ($\hat{a}_p\rightarrow\alpha_p$) with no depletion of this field over the relevant time scale of the system. Then, the interaction Hamiltonian in Eq.~(\ref{hamscat}) becomes
 \begin{equation}
\hat{H}_\text{int}=\hbar G_1\hat{a}^\dagger_+\hat{a}_\Omega+\hbar G_2\hat{a}_-^\dagger\hat{a}_\Omega^\dagger+H.c, \label{hameff}
\end{equation} 
with the enhanced EO couplings $G_{1/2}=\alpha_p g_\pm$. The first term in the Hamiltonian is responsible for coherent frequency conversion (CFC) between optical $\omega_+$ and microwave $\Omega$ photons and it has been extensively studied in Ref.~\cite{ilchenko_whispering-gallery-mode_2003, tsang_cavity_2011,Rueda:16,RuedaSanchez2018,rueda2019electrooptic}. The second term in Eq.~(\ref{hameff}) describes the parametric amplification and entanglement between optical $\omega_-$ and microwave photons $\Omega$ \cite{rueda2019electrooptic}. The entanglement between the two optical signals is achieved through their single interaction to the microwave mode. The entangled microwave photon created from the spontaneous parametric downconversion (SPDC), ruled by the second term in Eq.~(\ref{hameff}), is frequency up-converted to the optical mode $\omega_+$, ruled by the first term in Eq.~(\ref{hameff}), creating entanglement between the two optical modes.
\subsection{Resonant System Dynamics}
The basic building block of an electro-optic frequency comb is given by a pair of two optical modes coupled through a microwave mode. This systems can be achieved with the Stokes and anti-Stokes modes sharing the same pump mode as shown in Fig. \ref{setupsim}b, experimentally realized in Ref.~\cite{ilchenko_whispering-gallery-mode_2003}, or with a pair of dispersion-engineered isolated modes, where either the Stokes or anti-Stokes mode is suppresed as experimentally shown in~\cite{Rueda:16,willy2020,savchenkov_tunable_2009}. Assuming a general single port resonant systems with the waveguide coupling rates $\kappa_{e,k}$, internal loss rates $\kappa_{i,k}$, external  $\hat{a}^e_{k}$, and internal  $\hat{a}^i_{k}$ input noise terms for the participating modes $k$, the time evolution of the intra-cavity mode operators for the Hamiltonian introduced in Eq.~(\ref{hameff}) is given by the equations \cite{ilchenko_whispering-gallery-mode_2003,tsang_cavity_2011,rueda2019electrooptic}:
\begin{subequations}
\begin{align}
\dot{\hat{a}}_{+}&=-iG_1 \hat{a}_\Omega-\frac{\kappa_{+}}{2} \hat{a}_{+}+\hat{F}_+ ,    \label{rateeq_1} \\
\dot{\hat{a}}_{-}&=-iG_2 \hat{a}^\dagger_\Omega-\frac{\kappa_{-}}{2} \hat{a}_{-}+\hat{F}_- ,  \label{rateeq_2} \\
\dot{\hat{a}}_{\Omega}&= -iG^*_1 \hat{a}_+-iG_2 \hat{a}^\dagger_-  -\frac{\kappa_{\Omega}}{2} \hat{a}_{\Omega} + \hat{F}_\Omega.       \label{rateeq_3}
\end{align}
\end{subequations}
Where we have introduced $\kappa_k=\kappa_{e,k}+\kappa_{i,k}$  for total loss rate and $\hat{F}_k=\sqrt{\kappa_{i,k}} \hat{a}^i_{k}+\sqrt{\kappa_{e,k}} \hat{a}^e_{k}$ for the total input field operators of the mode $k$. $\hat{a}^j_{k}$ are zero mean quantum Langevin operators following the properties:
$\left[ \hat{a}_j(t),\hat{a}_j^\dagger(t')\right]= \delta(t-t')$ and $\langle\hat{a}_j(t) \hat{a}^\dagger_k(t')\rangle=(n_j(\omega,T) +1)\delta(t-t') \delta_{kj}$,
where $n_{j}=(\exp(\hbar\omega_j/(k_BT_j))-1)^{-1}$. The Eq.~(\ref{rateeq_1})-(\ref{rateeq_3}) are analytically solved in the Fourier domain to obtain the intra-cavity field operators $\hat{a}_k$. Furthermore, by using the single port input-output relations $\hat{a}_k^\text{out} = \sqrt{\kappa_{e,k}} \hat{a}_k  -\hat{a}^e_{k} $, we find the general solution:
\begin{equation}
\hat{\text{S}}^\text{out}(\omega)=\textbf{T}(\omega)\cdot \hat{\text{S}}^\text{in}(\omega), \label{FDC}
\end{equation}
where $\hat{\text{S}}^\text{out}(\omega)$ stands for the output field operators $[\hat{a}_+^\text{out}(\omega),\hat{a}_-^{\text{out}\dagger}(-\omega), \hat{a}_\Omega^{\text{out}}(\omega)]^\text{T}$, $\textbf{T}(\omega)$ stands for the transformation matrix, explicitly given in the supplemental material, and $\hat{\text{S}}^\text{in}(\omega)$ stands for the set of the input operators $[\hat{a}^e_{+},\hat{a}^i_{+},\hat{a}^{e\dagger}_{e,-},\hat{a}^{i\dagger}_{-}, \hat{a}^e_{\Omega},\hat{a}^i_{\Omega}]^\text{T}$. 
\subsection{Spontaneous Parametric Down-Conversion}
\begin{figure}[t]
	\centering
		\includegraphics[width=0.48\textwidth]{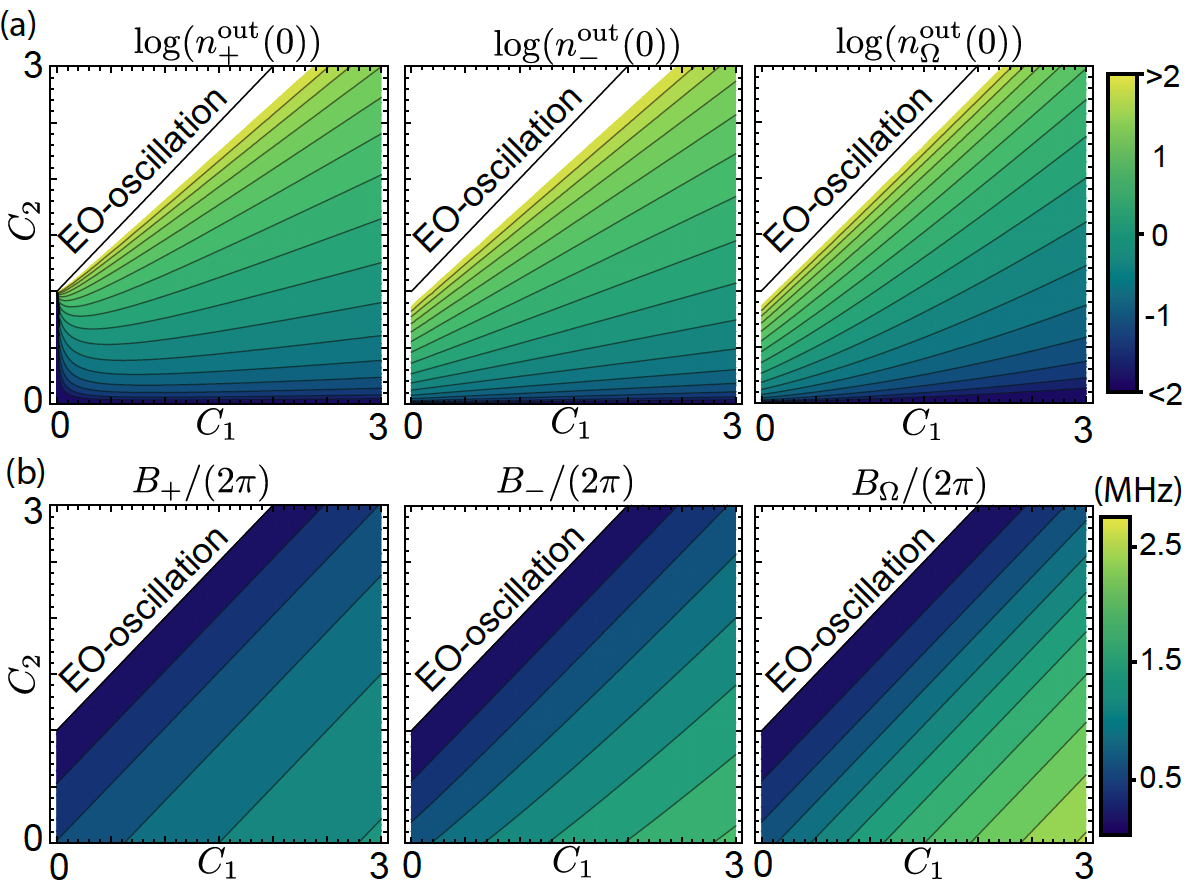}
	\caption{\textbf{(a)} Output photon generation at resonance for the anti-Stokes, Stokes and microwave modes as function of the multiphoton cooperativities $C_1$ and $C_2$ for $\eta_k=1$ and $\kappa_k=1$ MHz. We clearly see the generation of photon diverges to infinity for $C_2=1+C_1$ for all the modes.  The Stoke emission is always higher, even though the all modes share the same parameters. For $C_1=0$ the production rate follows the single electro-optic SPDC as presented in Ref.~\cite{rueda2019electrooptic}. \textbf{(b)} Generation bandwidth of the anti-Stokes, Stokes and microwave modes. The conversion bandwidth converges to zero at the divergence condition $C_2=1+C_1$ and for fixed values of $C2/C1$ the bandwidth remains nearly constant as shown in the straight contour lines. The modes show different bandwidth for the same cooperativities, being the anti-Stokes emission narrowest and the microwave emission broadest due to its direct coupling to both of the optical modes.}
	\label{maxbw}
\end{figure}
The output photon generation rate due to SPDC and CFC is given by $n_k^\text{out}(\omega)=\langle \hat{\text{a}}^{\text{out}\dagger}_k(\omega)\hat{\text{a}}_k^{\text{out}}(\omega) \rangle$:
\begin{subequations}
\begin{align}
n^\text{out}_+(\omega)&= 4\eta_+C_1C_2D(\omega),\label{SPDC3modes_1} \\
n^\text{out}_-(\omega)&= 4\eta_-C_2(1+C_1+4\omega^2/\kappa^2_+))D(\omega),\label{SPDC3modes_2} \\
n^\text{out}_{\Omega}(\omega)&= 4\eta_\Omega C_2(1+4\omega^2/\kappa^2_+)D(\omega) ,    \label{SPDC3modes_3}
\end{align}
\end{subequations}
where
\begin{eqnarray}
D^{-1}(\omega)&=\left(1+C_1-C_2-\frac{4\omega^2(\kappa_++\kappa_-+\kappa_\Omega)}{\kappa_+\kappa_-\kappa_\Omega}\right)^2&\nonumber\\
&+4\omega^2\left(\frac{1+C_1}{\kappa_-}+\frac{1-C_2}{\kappa_+}+\frac{\kappa_+\kappa_--4\omega^2}{\kappa_-\kappa_+\kappa_\Omega}\right)^2,&
\end{eqnarray}
$C_k=4G_k/(\kappa_k\kappa_\Omega)$ stands for the multiphoton cooperativity and $\eta_k=\kappa_{e,k}/\kappa_k$ are the waveguide coupling for the participating modes. The generated photons of the fields  $\hat{a}_+^\text{out}(\omega)$ and $\hat{a}_\Omega^\text{out}(\omega)$ are frequency anti-correlated  to $\hat{a}_-^\text{out}(-\omega)$ as expected from the energy conservation. The symmetry in the intra-cavity microwave $\Omega$ and optical $\omega_-$ photon numbers from the SPDC is broken due to the CFC into the anti-Stoke mode, leading to different output generation rates in amplitude and spectrum, as shown in Fig~\ref{maxbw}. 
At resonance ($\omega=0$) the pre-factor $D(0)^{-1}$  becomes $1+C_1-C_2$. Subsequently, $D(0)$ and the photon production diverges to infinity for  $C_2=1+C_1$. 
This marks the instability region of the system and becomes the threshold condition for the electro-optic parametric oscillation \cite{tsang_cavity_2011,RuedaSanchez2018,rueda2019electrooptic}, where the photon number increases exponentially in time until depleting the pump. Therefore, the linearization approximation in the Hamiltonian in Eq.~(\ref{hameff}) and (\ref{SPDC3modes_3})-(\ref{SPDC3modes_3}) breaks down and the analysis is not longer valid. Unlike the single electro-optic entanglement ($C_1\approx0$) described in Ref.~\cite{rueda2019electrooptic}, the bandwidth ($\text{B}_{k}$) of the generation output, usually defined as the FWHM of  $\hat{n}_k^\text{out}(\omega)$, is different for all the three modes even if all the modes share the same parameters $\kappa_k$. This effect can be clearly seen in Fig.~\ref{maxbw}b, where I show the emission bandwidth for different values of $C_k$. Furthermore, the bandwidth in all the modes follows a shifted quasi-linear behavior as function of $C_1\propto C_2$ and it converges to zero at the instability of $C_1-1=C_2$. The case of single microwave-to-optics entanglement can be found in the Eq.~(5)-(6) for $C_1=0$.

\subsection{Squeezed Quadratures}
The output state $\rho^\text{out}$  is a continuous variable (CV) three mode bosonic system described by a trace-class operator $\hat{\rho}$ (density matrix). To verify and quantify the quadrature squeezing, we define the dimensionless quadrature mode output operators $\hat{q}_k$ and $\hat{p}_k$ in terms of the output annihilation and creation operators given in Eq.~(\ref{FDC}) as:
\begin{equation}
\hat{q}_k=\frac{\hat{a}_k^\text{out}+\hat{a}_k^{\text{out}\dagger}}{\sqrt{2}}  \mbox{ and }\hat{p}_k= \frac{\hat{a}_k^\text{out}-\hat{a}_k^{\text{out}\dagger}}{\sqrt{2}i}\label{quad}.
\end{equation}
These operators follow $[q_k,p_k]=i$ and offer an analogy to the position and momentum operator of quantum harmonic oscillator\cite{Braunstein2005}. Moreover, the variances for the vacuum and coherent states are 0.5 according to the given definition in Eq.~(\ref{quad}). This ensemble $\rho^\text{out}$ is fully characterized by its first two statistical moments \cite{weedbrook2012}.  The mean value of the set of the six output quadratures $\hat{\textbf{x}}=\{\hat{q}_+,\hat{p}_+,\hat{q}_-,\hat{p}_-,\hat{q}_\Omega,\hat{p}_\Omega\}^\text{T}$  determines the first statistical moment of the system and for this case holds: $\langle\hat{\text{x}}\rangle=\text{Tr}(\hat{\rho}^\text{out} \hat{\text{x}})=0$. The second statistical moment corresponds to the covariance matrix (CM) and it is defined as:
\begin{equation}
V_{jk}=\frac{1}{2}\langle \Delta \hat{\text{x}}_j \Delta \hat{\text{x}}_k+\Delta \hat{\text{x}}_k\Delta \hat{\text{x}}_j\rangle\label{defcov}
\end{equation}
where $\Delta \hat{\text{x}}_j=\hat{\text{x}}_j-\langle \hat{\text{x}}_j\rangle$.  Applying a narrow filter function $\Theta(\omega)$ to the output fields around the resonance $\hat{\bar{a}}_k^\text{out}=\int\hat{a}_k^\text{out}(\omega)\Theta(\omega)\text{d}\omega$ and inserting into Eq.~(\ref{defcov}), the CM can be written as a function of the multiphoton cooperativities and the external (waveguide) $n^e_{\Omega}$ and internal $n^i_\Omega$ microwave noise occupation number:
\begin{widetext}
\begin{footnotesize}
\begin{equation}
\textbf{V} = \begin{bmatrix}
\left(0.5+\frac{4C_1(C_2+\bar{n}_\Omega)\eta_+}{(1+C_1-C_2)^2}\right)\textbf{I}&\left(\sqrt{4\eta_+\eta_- C_1C_2}\frac{1+C_1+C_2+2\bar{n}_\Omega}{(1+C_1-C_2)^2}\right)\textbf{Z} &\left(\sqrt{4\eta_+\eta_\Omega C_1}\frac{2C_2+2\bar{n}_\Omega-(1+C_1-C_2)n_\Omega^e}{(1+C_1-C_2)^2}\right)\textbf{I} \\
\left(\sqrt{4\eta_+\eta_- C_1C_2}\frac{1+C_1+C_2+2\bar{n}_\Omega}{(1+C_1-C_2)^2}\right)\textbf{Z} &\left(0.5+\frac{4C_2(C_1+1+\bar{n}_\Omega)\eta_- }{(1+C_1-C_2)^2}\right)\textbf{I} &\left(\sqrt{4\eta_-\eta_\Omega C_2}\frac{1+C_1+C_2+2\bar{n}_\Omega-(1+C_1-C_2)n_\Omega^e}{(1+C_1-C_2)^2}\right)\textbf{Z}  \\
\left(\sqrt{4\eta_+\eta_\Omega C_1}\frac{2C_2+2\bar{n}_\Omega-(1+C_1-C_2)n_\Omega^e}{(1+C_1-C_2)^2}\right)\textbf{I} &\left(\sqrt{4\eta_-\eta_\Omega C_2}\frac{1+C_1+C_2+2\bar{n}_\Omega-(1+C_1-C_2)n_\Omega^e}{(1+C_1-C_2)^2}\right)\textbf{Z}   &\left(0.5+\frac{4(C_2+\bar{n}_\Omega)\eta_\Omega }{(1+C_1-C_2)^2}+n^e_{\Omega}(1-\frac{4\eta_\Omega}{1+C_1-C_2})\right)\textbf{I} 
\end{bmatrix},
\end{equation}
\end{footnotesize}
\end{widetext}
where \textbf{I} is the identity matrix, \textbf{Z}=diag(1,-1) and $\bar{n}_\Omega=(\kappa_{e,\Omega} n^e_{\Omega}+\kappa_{i,\Omega} n^i_{\Omega} ) /\kappa_\Omega$ is the effective microwave thermal  mode number. The contribution of the optical thermal noise can be neglected due to their high frequency.  As expected from the emission behaviour, the CM diverges at the instability and all  covariances vanish when the SPDC coherent pump is off $C_2=0$ at 10 mK, leaving $\textbf{V}=0.5$\textbf{I} $_{6\times6}$.\\
The cross-quadrature squeezing is visualized using quasi-probability Wigner function in the phase space which is defined as:
\begin{equation}
W(\textbf{x})=\frac{\exp(-0.5\cdot\textbf{x}\cdot \textbf{V}^{-1}\cdot \textbf{x})}{\pi^2\sqrt{\text{det}[\textbf{V}]}}\label{wigner}.
\end{equation}
The normalized projection of the Wigner function for the cross quadratures $\{q_+,q_-\},\{q_-,q_\Omega\}$ and $\{q_+,q_\Omega\}$ in gray shades are shown in Fig.~\ref{wignerdelta}(a), where the blue lines highlight, where the projections reach their value $e^{-1}$ of their maximum for the parameters $C_1=0.5$, $C_2=0.9$ and $\eta_i=1$ at $T=0$. Similarly, the red line shows $e^{-1}$ values for the vacuum state $(C_2=0)$. 
We identify the relative quadrature squeezing $2\sqrt{2}\Delta q_-$ and $2\sqrt{2}\Delta q_+$ as the minor a major semiaxis of the blue highlighted ellipse depicted in the cross-correlation in Fig.~\ref{wignerdelta}(a), respectively. The quadrature squeezing in terms of the CM is given by:
\begin{figure}[t]
	\centering
		\includegraphics[width=0.50\textwidth]{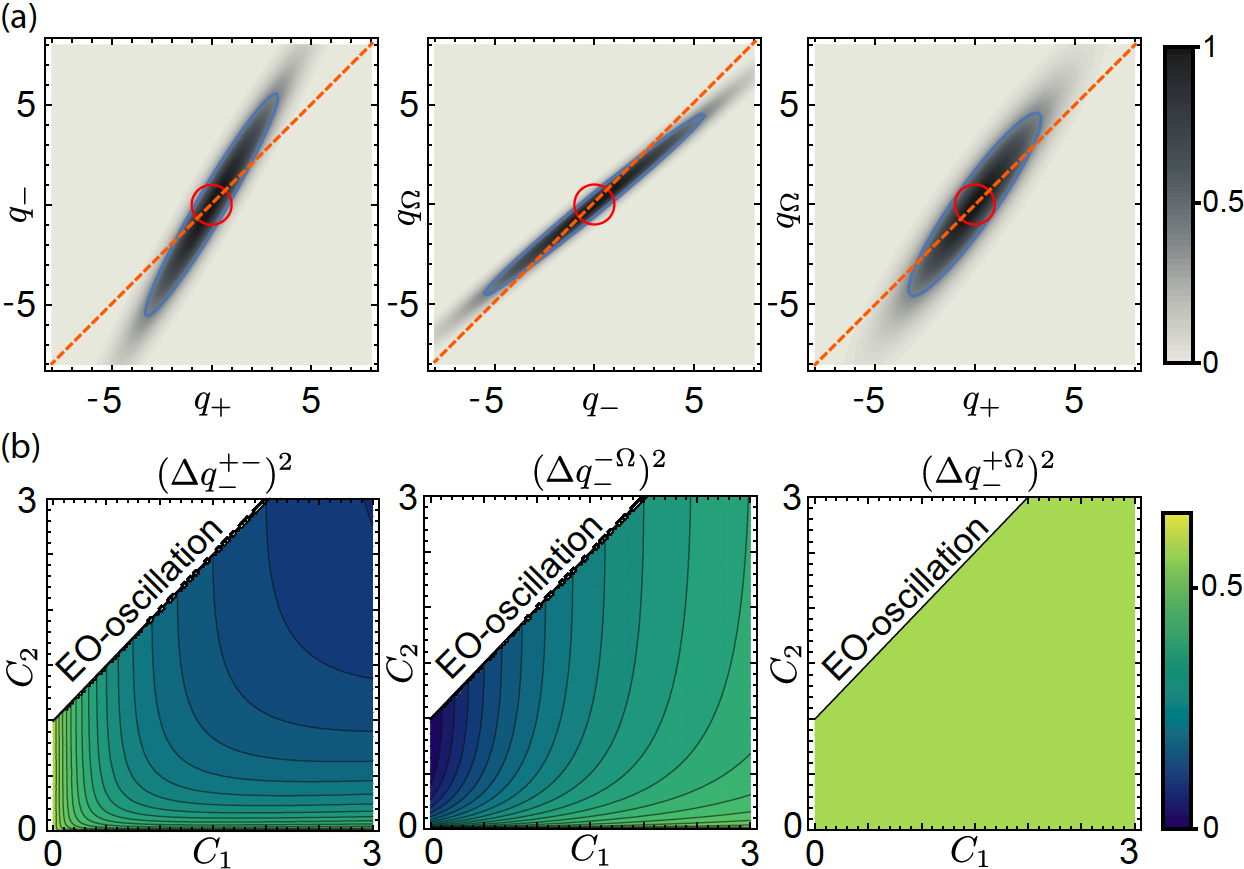}
	\caption{Cross-quadrature squeezing of the electro-optic output fields. \textbf{(a)} Normalized projections of the Wigner function of the three output quadrature pairs $\{q_+,q_-\}$, $\{q_-,q_\Omega\}$ and $\{q_+,q_\Omega\}$ for $\eta_k=1$, $\bar{n}_\Omega=0$, $C_1=0.5$ and $C_2=0.9$. The solid blue and red line indicates the drop by $e^{-1}$ of the maximum of the projection of the given Wigner function and the vacuum state, respectively. The blue elipses' semi-axis  are given by $2\sqrt{2}\Delta q_\pm$ and their  orientations by the squeezing angles $90^\circ-\Theta$, which do not coincide with $45^\circ$ of an ideal squeezer (dashed orange lines).  \textbf{(b)} Cross-quadrature squeezing as the function of the two multiphoton cooperativities $C_1,C_2$ for $\eta_k=1$. In general, for a fixed $C_2$ the value, the quadrature squeezing gets monotonically transduced to the anti-Stokes mode as a function of $C_1$.}
	\label{wignerdelta}
\end{figure}
\begin{small}
\begin{equation}
\Delta q^{lk}_{\mp}=\sqrt{\frac{ V_{ll}V_{kk}-V^2_{lk} }{V_{kk(ll)}\cos^2(\Theta)+V_{ll(kk)}\sin^2(\Theta)  \pm V_{lk} \sin(2\Theta)]}},\label{rsq}
\end{equation}\end{small}with  $k\ne i,j$ and the cross-quadratures'  angles following: $\tan(2\Theta)=\pm 2V_{lk}/(V_{ll}-V_{kk})$ \cite{rueda2019electrooptic}, which differ from the 45$^\circ$ for symmetric squeezers. In Fig.~\ref{wignerdelta}(a)  we observe the cross-quadrature two mode squeezing below the quantum limit in the diagonal directions for $\{q_+,q_-\}$ and $\{q_-,q_\Omega\}$. On the other hand, $\{q_+,q_\Omega\}$ shows no squeezing since the anti-Stoke and microwave mode are coherent but not entangled to each other. Furthermore, in Fig.~\ref{wignerdelta}(b) we show some values for $(\Delta q^{lk}_-)^2$ for the participating fields as function of the multiphoton cooperativities. For $C_1=0$ we observe in Fig.~\ref{wignerdelta}(b)  the strong  quadrature squeezing between microwave and the Stokes band as studied in \cite{rueda2019electrooptic}.  As  $C_1$ increases, the microwave radiation is coherently up-converted reducing $\Delta q^{-\Omega}_{-}$ and allowing the formation of  stronger than classical correlations between the anti- and Stokes modes. The purity of Gaussian state  ($\mathcal{P}=1/(2 \Delta q_-\Delta q_+)$ \cite{Paris2003}) is 1 when the variances of the quadratures fulfills the minimum uncertainty relation. For this system, $\Delta q^{+-}_\pm$  fulfill this condition when $\eta_\pm=1$ and $C_1\gg C_2\gg 1$, achieving also ideal EPR correlations  $q_+=q_-$ and $p_+=-p_-$.
\begin{figure}[t]
	\centering
		\includegraphics[width=0.50\textwidth]{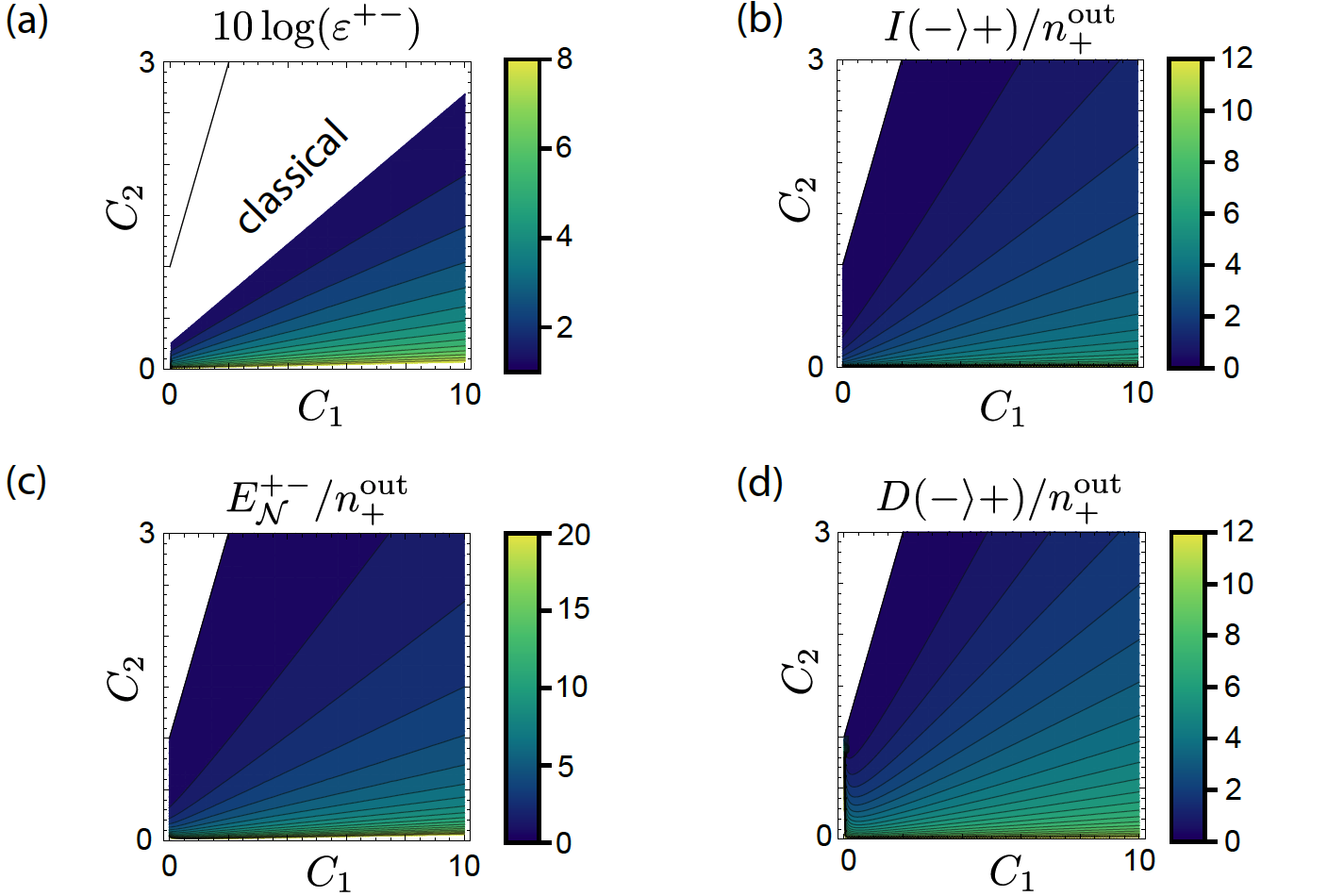}
	\caption{Entanglement metrics as function of the multiphoton cooperativities for $\bar{n}_\Omega=\eta_\Omega=0$ and $\eta_\pm=1$. \textbf{(a)} Schwarz inequality for the two output optical fields.  \textbf{(b)} Normalized coherent information $I(-\rangle+)/n^\text{out}_+$ setting the lower limit to the amount of qubits per Stokes photons. \textbf{(c)} Logarithmic negativity $E^{+-}_\mathcal{N}/n^\text{out}_+$ normalized by the output anti-Stokes emission, giving the higher limit of the number of ebits per anti-Stokes photons and \textbf{(d)} Normalized quantum discord $D(-\rangle+)/n^\text{out}_+$ setting the number of discordant bits per anti-Stoke photon. Top left white regions in the plots belong to the EO-Oscillation.}
	\label{wmetrics}
\end{figure}
\subsection{ Two Optical Modes Entanglement Metrics}
From the CM, stronger than classical correlation between the two optical modes are verified. From the Schwarz inequality, I define  $\varepsilon^{+-}=\log(|\langle a^\text{out}_+a^\text{out}_-\rangle|/\sqrt{n^\text{out}_+n_-^\text{out}})$, where $\varepsilon^{+-}\leq0$ is the limit for classical fields. In Fig.~\ref{wmetrics}(a) the region where the Schwarz inequality is violated $(\varepsilon^{+-}>0)$ is shown, proving the existence of entanglement between the two optical modes and defining a quantum channel.
Through standard metrics of quantum correlations, I evaluated the suitability of the two optical modes as a quantum channel.  The coherent information $I(-\rangle+)$ shows the lower bound for the distillable entanglement, i.e. the (asymptotic) amount of entangled bits  that can be extracted on average per copy of the state~\cite{Loyd97,weedbrook2012}. These values as a function of the cooperativities normalized by the emitted anti-Stokes photon number are shown in Fig.~\ref{wmetrics}(b) where the lower bound at the center of the plotted region $C_1=5$, $C_2=1.5$ gives an averaged estimation of 1.08 ebits per emitted anti-Stoke photons at the resonance ($I(-\rangle+)/n^\text{out}_+$). Similarly, through the logarithmic negativity $E^{+-}_\mathcal{N}$, which set the upper bound of the entangled ebits~\cite{vidal,weedbrook2012}, the maximum average per anti-Stokes photon at the center region (see Fig.~\ref{wmetrics}(c) ) is  $E^{+-}_\mathcal{N}/n^\text{out}_+=2.49$ ebits. In addition, the correlation between the two modes can be separated in a classical part and a quantum part. The quantum part is known as quantum discord~\cite{Olivares2012a,weedbrook2012}. The quantum correlation carried by each anti-Stoke photon is shown in Fig.~\ref{wmetrics}(d). It is important to note that  nonzero quantum discord does not necessarily mean entanglement, since it is not the only source of quantum correlation~\cite{Olivares2012a}.  Finally, this system does not provide a tripartite genuine entanglement source, since every witness or metric involving this requires $\langle \hat{a}_+^\text{out}(\omega) \hat{a}_-^\text{out}(\omega) \hat{a}_\Omega^\text{out}(\omega)\rangle\ne0$~\cite{Agustin2020}, which is not fulfilled by this system.
\section{Electro-optic frequency comb generation}
\begin{figure*}[t]
	\centering
		\includegraphics[width=1.0\textwidth]{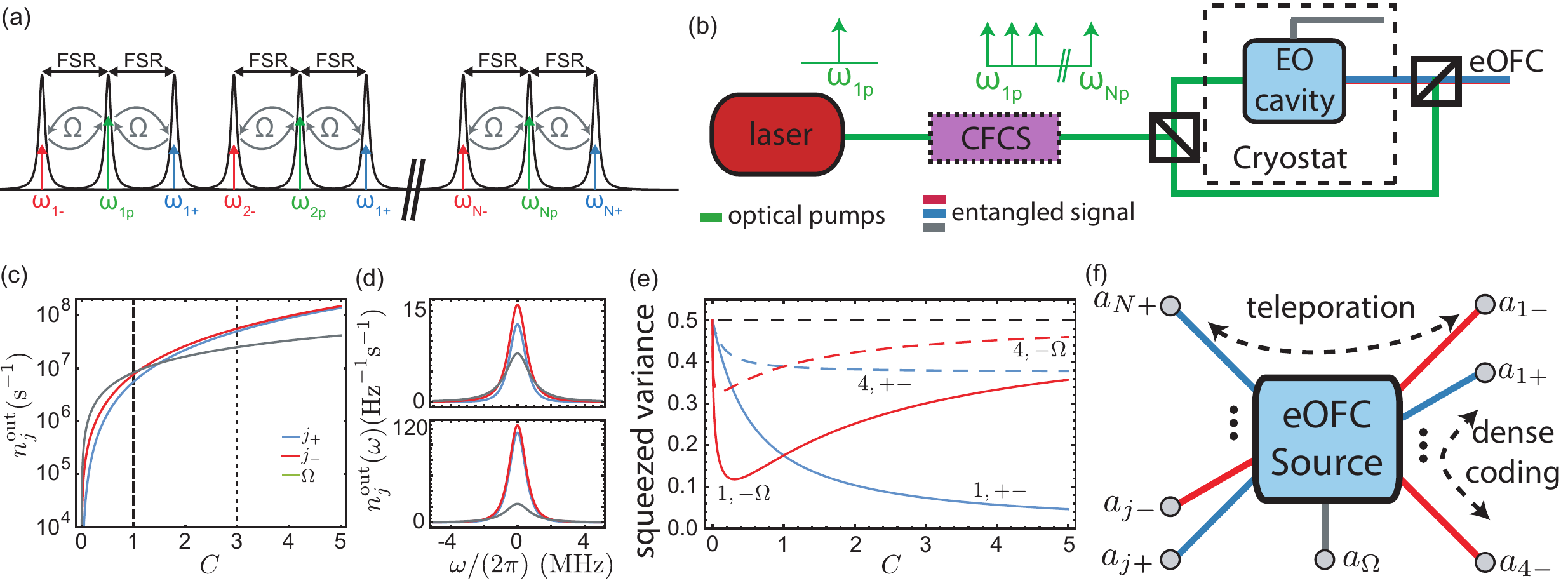}
	\caption{\textbf{(a)} An optical resonator with $\chi^{(2)}$ nonlinearity is pumped with $N$-coherent optical pumps on resonance separated 3 FSR from each other.  \textbf{(b)} A strong laser signal drives an electro-optic coherent frequency comb source as in \cite{Ruedacombs}, whose output consist of locked coherent signals spectrally separated by 3 FSR.  This scheme can be $N$-times repeated to add $2N$-lines to the comb. After the generation of the coherent pumps, each pump is split, with one arm sent to the system and the other is recombined at the cryostat output with the device's output total signal to suppress the pumps, leaving the entangled signal in the output. \textbf{(c)} Total photon production of the optical and microwave modes for $N=4$, $\eta_o=0.8$, $\eta_\Omega=0.5$, $\kappa_o=1.75$ MHz and $\kappa_\Omega=12.40$ MHz as function of the multiphoton cooperativity, for experimental feasible rates given in Ref.~\cite{Ruedacombs,willy2020} in an equally pumped system. For  $NC\gg1$ the optical photon emission in all the bands becomes equal. \textbf{(d)} Output photon generation spectrum of the optical modes $j_+$, $j_-$ and the microwave mode $\Omega$ at the $C=1$ (top) and $C=3$. The Bandwidths for the chosen values are  $B_+=1.11$, $B_-=1.20$ MHz and  $B_\Omega=1.72$ MHz. \textbf{(e)} Quadrature squeezing between the comb radiation for $N=1$ and $N=4$ with the paremeters $\eta_k=1$, $\bar{n}_\Omega=0$. \textbf{(f)} Electro-optic frequency multiplexed quantum channels. The Stokes modes are entangled to the anti-Stokes modes through the joint interaction with the microwave mode. This network offers playground for most of the protocols for CV quantum information.}
	\label{implementation}
\end{figure*}
\subsection{Device Implementation and Generation Scheme}
The proposed system for optical frequency combs generation is based on 3D-microwave cavity enclosing a mm-sized optical whispering gallery mode resonator~\cite{Rueda:16,Ruedacombs} operating at millikelvin temperatures as recently shown in Ref.~\cite{willy2020}. The optical FSR in those systems is in the range of the X-band and it remains almost constant over several FSR \cite{Ruedacombs}. The standard achievable optical quality factors for such systems are $Q_{i,o}\gtrsim10^8$ for lithium niobate~\cite{Rueda:16,Ruedacombs}  or $\gtrsim10^9$ for lithium tantalate~\cite{savchenkov_tunable_2009}, strongly decrease the needed optical pump power to achieve high number of optical intra-cavity photons. Furthermore, due to the high microwave confinement, these systems offer high (symmetric $g_+=g_-$) electro-optic coupling rates around $2\pi\times36.1$ Hz~\cite{willy2020}.

Electro-optic cooperativities approaching 1 has not yet been experimentally demonstrated. However,  It was measured in a cryogenic environment Ref.~\cite{willy2020}, that a mm-sized system has a relevant time scale for heating is in the order of one photon of noise per second for 1.5 mW  CW-pump ($\text{d}n^\text{noise}_\Omega/\text{d}t\text{d}P$=0.73 noise photons s$^{-1}$mW$^{-1}$). Therefore, in a scheme with an strong pulsed drive $\sim 0.1$ W, it can be easily achievable for time spans in the order of 1 $\mu$s without inducing internal  microwave noise. Moreover, by means of a cold waveguide, the effective microwave noise is reduced from the internal noise $n^i_\Omega$ as function of the waveguide coupling as $\bar{n}_\Omega=(1-\eta_\Omega)n^i_\Omega$. Time spans of 1 $\mu$s are more than enough for these devices to reach the steady state, since the device time scale is given by the spectral bandwidth as $\sim1/\text{B}$, which is in the orders of tens of nanoseconds \cite{Rueda:16,willy2020}. This allows stronger pump amplitudes or an increase of the repetition rate. Furthermore, a better WGM-resonator fabrication as in Ref.~\cite{Rueda:16,Ruedacombs,savchenkov_tunable_2009}  will reduce the needed optical pump power by a factor between $10^2$ to $10^4$ to reach $C>1$ in a system similar to Ref.~\cite{willy2020}.

In these systems the generation of two entangled optical fields spectrally separated by 2 FSR is achieved by driving strongly one mode of the resonator. This process could cascade to higher order modes similar to the microwave driven OFC coherent generation as in Ref.~\cite{Ruedacombs}. However, in a real system, the achievable occupation number of the first two sidebands won't be enough to drive the higher order modes and cascade. Furthermore, the threshold for considerable SPDC creation ($C\sim1$)  would set an experimentally  unreachable optical power to the main pump for a microwave noiseless frequency comb. 

A feasible $2N$-optical frequency comb with a single optical pump would require $N$ microwave modes in the cavity with frequencies of $n\cdot\text{FSR}$ and azimuthal spatial distributions of $2\cdot n$ to fulfill the energy and momentum conservation~\cite{Rueda:16}. In this case a given Stokes mode $\omega_p-n\cdot\text{FSR}$ is entangled only with the anti-Stokes mode $\omega_p+n\cdot\text{FSR}$ and the microwave mode  $n\cdot\text{FSR}$, restricting the exchange of information between different optical modes which is not ideal for quantum network.

A way to generate a $2N$ eOFC, useful for a quantum network, is through $N$ coherent optical pumps. The pump tones must be locked to the resonator's optical modes separated by 3 FSR or more. In this way the created entangled signals of two different pumps don't overlap as shown in Fig.~\ref{implementation}(a). A coherent optical frequency comb can be used as a multimode optical pump, achievable in a commercial electro-optic modulators or the more power efficient versions implemented in Refs.~\cite{Ruedacombs,loncarcomb}.  Furthermore, a coherent frequency comb with more controllable signal amplitudes can be built using commercial single-sideband suppressed-carrier electro-optic modulators (SSB). Where the laser beam is split into two arms and one goes through a SSB, creating a coherent customizable frequency shifted second pump, sharing the same intensity and phase fluctuations of the initial source. This method can be repeated $N$-times to generate the $N$ desirable pumps. Subsequently, by using notch filters, frequency wave multiplexer or arranging a Mach-Zehnder interferometer, at the end of the output in the cryostat as shown in Fig.~\ref{implementation}(b), the optical pumps can be filtered out from the optical frequency comb.

\subsection{2$N$-Optical Frequency Bin Generation}
For the case of 2$N$-optical modes, directly coupled to the same microwave mode through $g_{j\pm}$ and parametrically amplified by $\alpha_{j\pm}$, the general equations of motion for the fields' operators are given by:
\begin{subequations}
\begin{align}
\dot{\hat{a}}_{j+}&=-iG_{j+} \hat{a}_\Omega-\frac{\kappa_{j+}}{2} \hat{a}_{j+}+\hat{F}_{j+},& \label{manymodes1}   \\
\dot{\hat{a}}_{j-}&=-iG_{j-}\hat{a}^\dagger_\Omega-\frac{\kappa_{j-}}{2} \hat{a}_{j-}+\hat{F}_{j-} ,&    \label{manymodes2}\\ 
\dot{\hat{a}}_{\Omega}&= -i\sum\limits_1^N  (G_{j+}\hat{a}_{j+}+G^*_{j-} \hat{a}^\dagger_{j-})  -\frac{\kappa_{\Omega}}{2} \hat{a}_{\Omega}    +          \hat{F}_\Omega. &    \label{manymodes3}
\end{align} 
\end{subequations}
with $G_{j+}=\alpha_{j\pm}g_{j\pm} $. In the proposed system, the optical modes share the same parameters over several optical FSR~\cite{Ruedacombs}. Therefore, we assume $g_{j-}=g_{j+}$, $\kappa_{j\pm}=\kappa_o$ and $\eta_{j\pm}=\eta_j$. In addition, the pumps' tones are coherent to each other and they can be actively controlled in the system as discussed in the previous chapter. From this we can arrange a symmetric system with $G_{j\pm}=G$, allowing to find analytical solutions to the Eqs.~(\ref{manymodes1})-(\ref{manymodes3}) in the steady state. We find the photon emission of the SPDC given by:
\begin{subequations}
\begin{align}
n^\text{out}_{j+}&= 4\eta_o N C^2D_2(\omega) \\
n^\text{out}_{j-}&= 4\eta_oC(1+NC+4\omega^2/\kappa^2_o)D_2(\omega) \\
n^\text{out}_{\Omega}&= 4\eta_\Omega N C \left(1+4\omega^2/\kappa^2_o\right)D_2(\omega)     \label{threemodesDFG}
\end{align}
\end{subequations}
where $D_2^{-1}(\omega)=\left(1+\frac{4\omega^2}{\kappa^2_o}\right)^2\left(1+\frac{4\omega^2}{\kappa^2_\Omega}\right).$
From the solutions we can find that the generation of the $2N$-optical sidebands due to SPDC are $N$-times amplified in each mode, which leads to a grow of $\sim N^2$ for the total SPDC photon generation. 
Due to the EO-coupling symmetry between the anti-Stokes and Stokes modes, the system never achieves the EO-oscillations region and the total production and the emission spectrum of the optical modes becomes equal as the multiphoton cooperativity increases ($NC^2\gg C$) as shown in Fig.~\ref{implementation}(c)-(d).
\subsection{Covariance Matrix and Quadrature Squeezing}
Similarly to the previous section, I build the covariance matrix by amplifying a narrow filter to the output fields and calculate correlation of the quadrature of each comb line. This leads to the quadratic $2N+1$ covariance matrix:
\begin{equation}
\textbf{V}=
\begin{bmatrix}
\text{V}_{+}\textbf{I}& \text{V}_{+-}  \textbf{Z} & \text{V}_{++} \textbf{I}   &       \text{V}_{+-}\textbf{Z}     &  \cdots     & \text{V}_{+\Omega}    \textbf{I}  \\
\text{V}_{+-}\textbf{Z} & \text{V}_{-}\textbf{I}& \text{V}_{+,-}\textbf{Z} & \text{V}_{--}\textbf{I} &      \cdots   &  \text{V}_{-\Omega}    \textbf{Z} \\
\text{V}_{++} \textbf{I}& \text{V}_{+-}\textbf{Z} & \ddots & \ddots & \ddots &  \vdots   \\
   \text{V}_{+-}\textbf{Z} & \text{V}_{--}\textbf{I}& \ddots & \ddots & \ddots &  \text{V}_{+\Omega}    \textbf{I} \\
 \vdots& \ddots        & \vdots & \ddots &  \text{V}_{-}\textbf{I}&  \text{V}_{-\Omega}\textbf{Z}  \\
\text{V}_{+\Omega}\textbf{I}& \text{V}_{-\Omega}\textbf{Z}     &    \cdots     &  \text{V}_{+\Omega}    \textbf{I} &  \text{V}_{-\Omega}\textbf{Z}       & \text{V}_{\Omega}\textbf{I}
\end{bmatrix}
\label{matrixfull}
\end{equation}
with the CV-elements explicitly given as:
\begin{subequations}
\begin{align}
\text{V}_{+}&= 0.5+4\eta_oC(\bar{n}_\Omega+NC)\\
\text{V}_{-}&= 0.5+4\eta_oC(\bar{n}_\Omega+1+NC)\\
\text{V}_{\Omega}&= 0.5+4\eta_\Omega(\bar{n}_\Omega+NC)+n^e_\Omega(1-4\eta_\Omega)\\
\text{V}_{+-}&= 2\eta_oC(2\bar{n}_\Omega+1+2NC) \\
\text{V}_{+\Omega}&= \sqrt{4\eta_o\eta_\Omega C}(2\bar{n}_\Omega-n^e_\Omega+2NC)\\
\text{V}_{-\Omega}&=  \sqrt{4\eta_o\eta_\Omega C}(2\bar{n}_\Omega-n^e_\Omega+1+2NC)\\
\text{V}_{++}&=  4\eta_o C(\bar{n}_\Omega+ NC)\\
\text{V}_{--}&= 4\eta_o C(\bar{n}_\Omega+ 1+NC).\label{matrixchanges}
\end{align}
\end{subequations}
\begin{figure*}[t]
	\centering
		\includegraphics[width=1.0\textwidth]{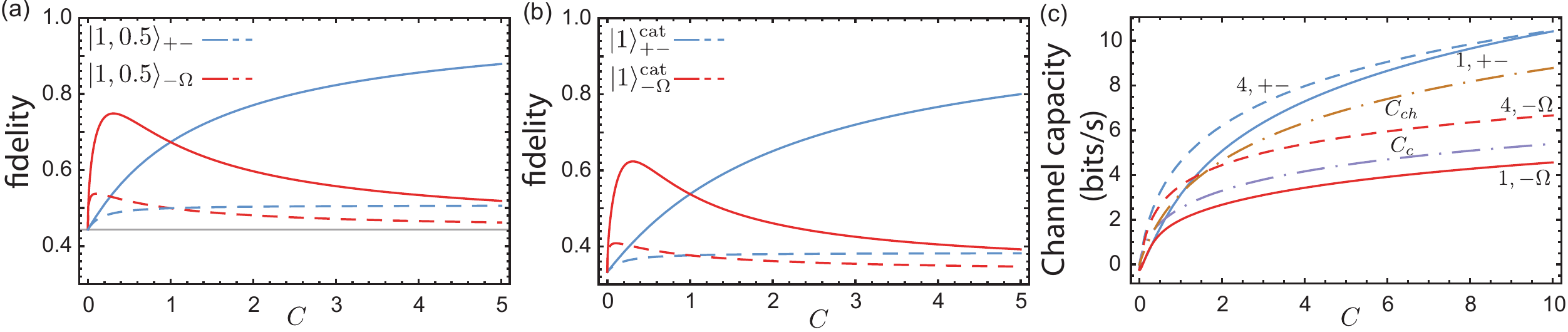}
	\caption{Quantum network channels. \textbf{(a)} Teleportation fidelities for a coherent squeezed state (Gaussian state). The fidelity for the $N=1$ (solid lines) between the Stoke and anti-Stokes mode converges to 1 for $C\gg1$. The Stokes and microwave modes achieves their maximum fidelity for cooperativities below the unity and it converges to its classical limit (gray line) for high $C$. The teleportation classical limit for fidelity is given by 0.5 in case of coherent states and gets reduced as the input state squeezing $F_{cl}=e^{-r}/(1+e^{2r})$~\cite{Owari_2008}. \textbf{(b)} Fidelity values for a cat state (non-Gaussian state). For the case $N=1$ the $F^{+-}$ also achieve unity for $C\gg1$ but the $F^{-\Omega}$ get a lower higher bound. In \textbf{(a)} and \textbf{(b)} the dashed lines represent the fidelities for a $N=4$ network, where the achievable fidelities are still above the classical limit but its maximum value get strongly reduced as expected from the non-cloning theorem.
	\textbf{(c)} Unconditional dense coding capacity for the channels anti-Stokes to Stokes and Stokes to microwave for $N=1$ (solid lines) and $N=4$ (dashed lines) as a function of the multiphoton cooperativity. The classical capacity of a quantum channel is increased by sharing an entangled state. The capacities for a coherent state single $C_{c}$ and dual quadrature encoding $C_{ch}$ for $N=1$ (dashed point lines) for the same amount of emitted photon are surpassed by the created channels from the eOFC.}
	\label{squeezing}
\end{figure*}
In this system the cross quadrature squeezing appears between the anti- and Stokes modes $\{q_{j-},q_{j+}\}$ and the Stokes-microwave $\{q_{j-},q_\Omega\}$ modes. The squeezing and anti-squeezing for these combinations are given as:
\begin{subequations}
\begin{align}
\Delta q^{N,-+}_\pm&= \sqrt{\frac{0.5+4\eta_oC+8\eta_oC^2(N-\eta_o)}{1+4\eta_oC(1+2NC\mp\sqrt{1+(1+2NC)^2})} }\label{combsqueezing1}\\
\Delta q^{N,-\Omega}_\pm&= \sqrt{ \frac{0.5+\Psi-8C\eta_\Omega\eta_o}{1+\Psi\mp\sqrt{\Psi^2+16\eta_o\eta_\Omega C}}}\label{combsqueezing2},
\end{align}
\end{subequations}
with $\Psi=4C(\eta_o(NC+1)+N\eta_\Omega)$. A remarkable feature follows that for $NC\gg1$, $(\Delta q^{N,-+}_-)^2$ achieves its minimum value of $0.5(1-\eta_o/N)$, which means that the variance squeezing is distributed equally between the $N$ modes. On the other hand, the Stokes-microwave squeezing  converges back to the vacuum limit $\Delta q^{N,-+}_-=\sqrt{0.5}$ after achieving its minima as shown in Fig.~\ref{implementation}(e).
\section{Electro-optic quantum network}

The entangled frequency comb signals are indeed bosonic quantum channels and they can be used as a source for a quantum network, where each anti- and Stokes modes works as a quantum channel at different frequencies as shown in Fig.~\ref{squeezing}. In this system each Stokes channel is coherent to the other Stokes channels and entangled to the microwave mode and to all the anti-Stokes channels.

To test the suitability of the generated quantum channels, we take the example of a Gaussian quantum protocol as teleportation. Assuming a Braunstein-Kimble scheme,  Alice (sender) combines an unknown quantum state with one arm of the entanglement source with a beam splitter and performs the corresponding Bell measurements on the output $x_-$, $p_+$ . Then, this information is sent classically to Bob (receiver), where an appropriate phase space displacement in the second arm of the entanglement source is performed to complete the state transfer \cite{braunsteinkimble}. The figure of merit for teleportation is given by the fidelity, defined as the overlap between input and output state $F= \pi\int W_\text{in}(\beta)W_\text{out}(\beta)\text{d}^2\beta$.  The teleportation fidelity of a Gaussian and non-Gaussian input state, such as a squeezed coherent  $|\psi_\text{in}\rangle=|\alpha=1,r=0.5\rangle$ and a Yurke-Stoler cat state $|\psi_\text{in}\rangle=(|\alpha=1\rangle-i|-\alpha=-1\rangle)/\sqrt{2}$ are calculated from the squeezing as given in Ref.~\cite{rueda2019electrooptic}, where Alice is at the channel $a_{j-}$  and Bob at $a_{j_+}$ or $a_{\Omega}$. In Fig.~\ref{squeezing}(a) the teleportation fidelity for a quantum network for the symmetric case $C_1=C_2$  for the over-coupled limit $\eta_k=1$ is shown in Fig.~\ref{squeezing}(a)-(b). The solid lines in both figures show the values for a $N=1$ network and the dashed lines for a $N=4$. The fidelity for the the coherent squeezed state and the cat state are shown in in Fig.~\ref{squeezing}(a) and (b), respectively.  For both state types, $F^{+-}$ goes to unity and $F^{-\Omega}$ goes to its classical limit for $C\gg1$ and  $N=1$. On the other hand, the $N=4$ symmetric network shown in dashed lines in Fig.~\ref{squeezing}(a)-(b), the fidelity is bounded below 1 for the four channels Stokes and anti-Stokes channels due to the no-cloning theorem \cite{noncloning}. In this case, all the receiving parties at the anti-stokes modes have to cooperate and share the information of their projective measurements to the channel where the states have to be reconstructed as discussed in \cite{Lockbraunstein}.

The classical channel capacity can be enhanced by sharing an entangled state~\cite{bennett92}.  This increase is commonly known as quantum dense coding. In this case one arm of the entangled state is sent to bob and the other is sent to Alice, who perform a phase space translation $D(\alpha)$ (modulation) and sends the output signal to Bob, who retrieves the information by combining the entangled arm Alice signal in a 50:50 beam splitter and followed by Bells measurements. By using this method the rate of the classical capacity can be exceed depending on the level of squeezing as studied in Ref.\cite{Densecodingbraunstein}. The capacity of the channel can be estimated as given in Ref.~\cite{Ralph2002} and the results are given in Fig.~\ref{squeezing}(c) for ideal Bell-measurements for networks of $N=1$ and $N=4$. The photon number and the quadrature squeezing scale differently with increasing cooperativity as shown in Fig.\ref{implementation}(c)-(e). Therefore, I compare the capacities for coherent single and dual encoding quadrature, whose photon number increase with $n^\text{out}_-(0)$, for reference.  In Fig.\ref{squeezing}(c) single and dual encoding are surpassed by the created anti-Stokes to Stokes channel of $N=1$ and $N=4$.

\section{Conclusions}

I have presented the theoretical treatment for the generation of multimode electro-optic entangled channels through SPDC and CFC. I studied the generation in the case of two optical modes, identifying the stability regions and  showing the performance of the system as quadrature squeezer and entangler, where important experimental parameter such as asymmetric coupling rates and microwave added noise number were handled and built into the analytical results. Further, the study is extended to $2N$ optical modes, where I propose a specific pulsed driving mode of a system already built and tested at the needed cryogenic temperature~\cite{willy2020}, whose optical enhancement has been already proven in similar setups~\cite{Rueda:16,savchenkov_tunable_2009}. This makes feasible high multiphoton electro-optic cooperativities for period of time vastly surpassing the system time scales, reaching the steady state. Furthermore, the required coherent pump needed to drive the system is based on commercial devices.

I studied the performance of the device as a quantum network by analyzing the protocols of unconditional quantum state teleportation and dense coding between the channels (modes). The results verify the higher capabilities of this network in comparison to classical one by profiting from multimode quadrature squeezing, justifying its experimental test and further developments. 

The presented system is a versatile source, which for  $N=1$ and $C_1=0$ can be used for experiments in electro-optic quantum illumination \cite{sihui2008,Barzanjeh2015,tsang_cavity_2011} and for $N>1$ produces multi-photon entangled quantum states similar to its Kerr-combs counterpart \cite{Reimer1176,kues2017}, which finds uses in  research areas such as optical quantum computation~\cite{menicci2008}, optical sensing beyond the classical limit~\cite{kolobov99} and high dimensional quantum states~\cite{cozzolino19}.


\section{Acknowledgements}
I thank Prof. Shabir Barzanjeh and Dr. Ulrich Vogl  for the fruitful discussions. 
\section{Authors contribution}
 Analytical model and analysis was done by A.R. 
\section{Additional information}
The numerical data generated in this work is available from the author upon reasonable request. Supplementary information is available online.

\bibliography{FinkGroupBib_v7}
\bibliographystyle{ieeetr}

\newpage
\onecolumngrid
\newpage
 
\renewcommand{\thefigure}{S\arabic{figure}}
\appendix
\tableofcontents
\setcounter{figure}{0}

\section{Two Opticals Modes}
\subsection{Hamiltonian and Equation of Motion}
The electro-optic interaction can be described by using two main modes, one in the microwave $\Omega$ and one in the optical $\omega_o$ regime, which interact in a nonlinear material with second order polarizability response $\vec{P}=\chi^{(2)}$. This nonlinear response, originates two new oscillation in the dipoles of the system with frequencies $\omega_\pm=\omega_0\pm\Omega$, creating the so called Stokes and anti-Stokes modes. To describe this system the full Hamiltonian can be written as :
\begin{equation}
\hat{H}=\hbar\omega_o\hat{a}_o+\hbar\omega_+\hat{a}_++\hbar\omega_-\hat{a}_-+\hbar\omega_\Omega\hat{a}_\Omega+H_I
\end{equation}
where $H_I$ is the electro-optic interaction hamiltonian given by~\cite{RuedaSanchez2018}:
\begin{equation}
H_I=\hbar g_-\hat{a}_-^\dagger\hat{a}_\Omega^\dagger\hat{a}_o+\hbar g_+\hat{a}_+^\dagger\hat{a}_\Omega\hat{a}_o+\text{H.c.} \label{4modeha}
\end{equation}
where $\hat{a}$, $\hat{a}_\pm$ and $\hat{a}_\Omega$ are the optical pump, optical anti-Stokes, Stokes and microwave mode annihilation operators. The first term in the Hamiltonian is the so called difference frequency generation (DFG), known as the Stokes band, and the second term is responsible for the frequency sum generation (SFG),  known as the Anti-stokes band. The coupling rate, which determines the creation rates per pump photon is given by the mode overlap and the electro-optic coefficient as: 
\begin{equation}
g_\pm=-\frac{n_pn_\pm}{n_\Omega}r \sqrt{\frac{\hbar \omega_p\omega_\Omega\omega_\pm}{8\epsilon_0 V_pV_\Omega V_\pm}} \int dV \psi_\pm^* \psi^{(*)}_\Omega\psi_p,\label{eq:defgg}
\end{equation}
where $\psi_k$ are the spatial mode distribution, $V_k$ are the effective mode volume, $n_k$ are the refractive index and  $\omega_k$ are the angular frequency.
The evolution of the operators using the Heisenberg picture leads to the coupled equations:
\begin{subequations}
\begin{align}
\dot{\hat{a}}_o&=-i\omega_o\hat{a}-ig_-^*\hat{a}_{-}\hat{a}_\Omega+-ig_+^*\hat{a}_\Omega^\dagger\hat{a}_+, \label{pa}\\
\dot{\hat{a}}_+&=-i\omega_+\hat{a}_+-ig_+\hat{a}_\Omega\hat{a}_{o},\\
\dot{\hat{a}}_-&=-i\omega_-\hat{a}_--ig_-\hat{a}_\Omega^\dagger\hat{a}_{o},\\
\dot{\hat{a}}_\Omega&=-i\Omega\hat{a}_\Omega-ig_-\hat{a}^\dagger_{-}\hat{a}_{o}-ig_+^*\hat{a}_{o}^\dagger\hat{a}_+.
\end{align} \label{ratesimple}
\end{subequations}
The system is driven with a strong coherent pump $\omega_p=\omega_o$, whose contribution is many order of magnitudes higher than the last two terms Eq.~(\ref{pa}). Therefore, $a_o$ is treated as a complex number $\alpha_p=\langle \hat{a}_o\rangle$. I handle an open system coupled to waveguide and to an internal bath through a rate $\kappa_{e,k}$ and $\kappa_{i,k}$, respectively. Assuming that $\omega_\pm=\omega_o+\text{FSR}$  we have.
\begin{subequations}
\begin{align}
\dot{\hat{a}}_+&=-\kappa_+\hat{a}_+-i\alpha_pg_+\hat{a}_\Omega+ \sqrt{\kappa_{e,+}}\hat{a}^e_++\sqrt{\kappa_{i,+}}\hat{a}^i_+,  \\
\dot{\hat{a}}_-&=-\kappa_-\hat{a}_--i\alpha_pg_-\hat{a}_\Omega^\dagger+\sqrt{\kappa_{e,-}}\hat{a}^e_-+\sqrt{\kappa_{i,-}}\hat{a}^i_-,\\
\dot{\hat{a}}_\Omega&=-\kappa_\Omega\hat{a}_\Omega-i\alpha_pg_-\hat{a}^\dagger_{-}-i\alpha_p^*g_+\hat{a}_+ +\sqrt{\kappa_{e,\Omega}}\hat{a}^e_\Omega+\sqrt{\kappa_{i,\Omega}}\hat{a}^i_\Omega.
\end{align} \label{ratesimpletres}
\end{subequations}
where I have introduce the zero-mean input optical and microwave input losses which follows the relations:
\begin{subequations}
\begin{align}
\left[ \hat{a}_j(t),\hat{a}_j^\dagger(t')\right]&= \delta(t-t') \label{commu}\\
\langle\hat{a}_j(t) \hat{a}^\dagger_k(t')\rangle&=(n_j(\omega,T) +1)\delta(t-t') \delta_{kj}\label{noisen},
\end{align}
\end{subequations}
\subsection{Fourier analysis and Transformation matrix}
The asymmetric system can also be seen mathematically as separated coherently pumped two-optical mode system, where either $a_+$ or $a_-$ is set to zero, as shown in Fig.~\ref{setupsim}c in the main text. By moving to the Fourier domain to obtain the microwave and optical resonator fields and then, substituting the solutions
of Eqs.~\ref{ratesimpletres} into the single port  input-output relations:
\begin{equation}
\hat{a}^\text{out}_k=-\hat{a}^\text{in}_k +\sqrt{\kappa_{e,k}}\hat{a}_k
\end{equation}
we have the following solutions to the equation:
\begin{subequations}
\begin{align}
\hat{a}^\text{out}_+(\omega)&=T_{11}(\omega) \hat{a}_{e,+}(\omega)+T_{12}(\omega) \hat{a}_{i,+}(\omega)+T_{13} (\omega)\hat{a}^\dagger_{e,-}(\omega)  +T_{14} (\omega)\hat{a}^\dagger_{i,-}(\omega)   +T_{15} (\omega)\hat{a}_{e,\Omega} (\omega)  +T_{16} (\omega)\hat{a}_{i,\Omega} (\omega)    \\
\hat{a}^\text{out}_-(\omega)&=T^*_{21}(\omega) \hat{a}^\dagger_{e,+}(\omega)+T^*_{22} (\omega)\hat{a}^\dagger_{i,+}(\omega)+T^*_{23}(\omega) \hat{a}_{e,-}(\omega)  +T^*_{24}(\omega) \hat{a}_{i,-}(\omega)   +T^*_{25}(\omega) \hat{a}^\dagger_{e,\Omega} (\omega)  +T^*_{26} (\omega)\hat{a}^\dagger_{i,\Omega}  (\omega)   \\
\hat{a}^\text{out}_\Omega(\omega)&=T_{31}(\omega) \hat{a}_{e,+}(\omega)+T_{32}(\omega) \hat{a}_{i,+}(\omega)+T_{33}(\omega) \hat{a}^\dagger_{e,-} (\omega) +T_{34} (\omega)\hat{a}^\dagger_{i,-} (\omega)  +T_{35} (\omega)\hat{a}_{e,\Omega}(\omega)   +T_{36}(\omega) \hat{a}_{i,\Omega} (\omega)   
\end{align} \label{ratesimple}
\end{subequations}
where the coefficients $T_{jk}$ can be written in a matrix as follow:
\begin{footnotesize}
\begin{equation}
\textbf{T} = \frac{1}{\text{M}}\begin{bmatrix}
(\Gamma_-^*\Gamma_\Omega-G_2^2)\gamma^{e+}_{e+}-\text{M}&(\Gamma_-^*\Gamma_\Omega-G_2^2)   \gamma^{e+}_{i+}&-G_1G_2\gamma^{e+}_{e-}& -iG_1G_2 \gamma^{e+}_{i-} &-iG_1\Gamma^*_-\gamma^{e+}_{e\Omega} & -iG_1\Gamma_-^*\gamma^{e+}_{i\Omega} \\
G^*_1G^*_2\gamma^{e-}_{e+}& G^*_1G^*_2      \gamma^{e-}_{i+}&(\Gamma_+\Gamma_\Omega+G_1^{*2})\gamma^{e-}_{e-}-\text{M} &(\Gamma_+\Gamma_\Omega+G_1^{*2})\gamma^{e-}_{i-}  &iG_2\Gamma_+\gamma^{e-}_{e\Omega} & iG_1\Gamma_+ \gamma^{e-}_{i\Omega}\\
-iG_1^*\Gamma_-^*\gamma^{e\Omega}_{e+}& -iG_1^*\Gamma_-^* \gamma^{e\Omega}_{i+}& -iG_2\Gamma_+\gamma^{e\Omega}_{e-} &  -iG_2\Gamma_+\gamma^{e\Omega}_{i-}&\Gamma_+\Gamma_-^*\gamma^{e\Omega}_{e\Omega}-\text{M}& \Gamma_+\Gamma_-^*\gamma^{e,\Omega}_{i\Omega} \label{matrix}
\end{bmatrix},
\end{equation}
\end{footnotesize}
with $\text{M}(\omega)=\Gamma_+\Gamma^*_-\Gamma_\Omega+|G_1|^2\Gamma_-^*-|G_2|^2\Gamma_+$, $\gamma^{mn}_{jk}=\sqrt{\kappa_{m,n}\kappa_{j,k}}$ and $\Gamma_k=\kappa_k/2-i\omega$. 
As in the main text in the input fields can be written as:
\begin{equation}
\hat{\text{S}}^\text{out}(\omega)=\textbf{T}(\omega)\cdot \hat{\text{S}}^\text{in}(\omega), \label{2twomodessupp}
\end{equation}
where $\hat{\text{S}}^\text{out}(\omega)=[\hat{a}_+^\text{out}(\omega),\hat{a}_-^{\text{out}\dagger}(-\omega), \hat{a}_\Omega^{\text{out}}(\omega)]^\text{T}$ and $\hat{\text{S}}^\text{in}(\omega)=[\hat{a}^e_{+}(\omega),\hat{a}^i_{+}(\omega),\hat{a}^{e\dagger}_{-}(-\omega),\hat{a}^{i\dagger}_{-}(-\omega), \hat{a}^e_{\Omega}(\omega),\hat{a}^i_{\Omega}(\omega)]^\text{T}$. \\
The covariance matrix is then written using Eq.(\ref{quad}) and (\ref{defcov}) given in the main text. For the following sections, we defined the reduce covariance matrix (two modes) 
\begin{equation}
\text{V}^{jk} = \begin{bmatrix}
V^{jk}_11\textbf{I} &V^{jk}_{13}\textbf{Z}  \\
V^{jk}_{13}\textbf{Z}   &V^{jk}_33\textbf{I} 
\end{bmatrix},
\end{equation}
with $V^{jk}_{11}=n^\text{out}_j+0.5$, $V^{jk}_{33}=n^\text{out}_k+0.5$ and $V^{jk}_{13}=\langle \hat{q}_j\hat{q}_k+\hat{q}_k\hat{q}_j \rangle/2$ and where $j,k\in{\{+,-,\Omega\}}$.


\subsection{Entanglement metrics}

\subsubsection{Cauchy-Schwarz Criterion}

Entanglement is a pure quantum effect. Therefore, we expect that the joint state between the output microwave and optical field does not have  a proper $P$-representation and it violates the inequality
\begin{equation}
|\langle\hat{a}_j^\text{out}\hat{a}_k^\text{out}\rangle|\le\sqrt{n^\text{out}_j n^\text{out}_k}
\end{equation}
This is the Cauchy-Schwarz inequality which is one criterion to separate quantum and classical fields. This phase sensitive cross correlation remains under this bound for classical states and it can be violated by a field with a $P$-function with a negative region. Therefore, one common metric for entanglement is given as:
\begin{equation}
\varepsilon\equiv\log\left|\frac{\langle\hat{a}_j^\text{out}\hat{a}_k^\text{out}\rangle}{   \sqrt{n^\text{out}_j n^\text{out}_k} }\right| \label{inequality}
\end{equation}
For fields fulfilling  $\varepsilon>0$ we say they are entangled and they belong to a quantum state and for $\varepsilon\le0$ we say that they are classical fields. For the electrooptic system, we get the following expressions:
\begin{eqnarray}
\varepsilon_{+-}&=&\log\left( \frac{1+C_1+C_2+2\bar{n}_\Omega}{\sqrt{4(C_2+\bar{n}_\Omega)(C_1+1+\bar{n}_\Omega)}}\right),\\
\varepsilon_{-\Omega}&=&\log\left( \frac{1+C_1+C_2+2\bar{n}_\Omega-(1+C_1-C_2)n_\Omega^e}{\sqrt{  (C_1+1+\bar{n}_\Omega)( 4(C_2+\bar{n}_\Omega) +n_\Omega^e((1+C_1-C_2)^2 /\eta_\Omega -4 (1+C_1-C_2))  }}\right),\\
\varepsilon_{+\Omega}&=&\log\left( \frac{2C_2+2\bar{n}_\Omega-(1+C_1-C_2)n_\Omega^e}{\sqrt{  (C_2+\bar{n}_\Omega)( 4(C_2+\bar{n}_\Omega) +n_\Omega^e((1+C_1-C_2)^2 /\eta_\Omega -4 (1+C_1-C_2))  }}\right)\leq0.
\end{eqnarray}
Then, the condition for the output  "quantum" fields depends exclusively on the cooperativity and the added microwave thermal noise. This metric offers the same results between $\varepsilon_{+-}$ and $\varepsilon_{-\Omega}$ for a waveguide in ground state $n^e_\Omega=0$ ensuring entanglement between the two optical fields and the Stokes-microwave entanglement. It is important to point out that $\varepsilon_{+\Omega}$ has its upper bound at 0 ruling out entanglement between these two modes. 

\subsubsection{Coherent Information}

Coherent information  provides the lower bound to the number of ebits which can be distillable from the source and its numerical value is defined  as:
 \begin{equation}
I(\rho_j\rangle\rho_k)=S(\rho_k)-S(\rho_{j,k})
\end{equation}
where $S(\rho_k)$ is the von Neumann entropy of the output state in the mode $k$ and $S(\rho_{j,k})$ is the  joint entropy of the output state. For Gaussian states the coherent information is calculated from the symplectic eigenvalues $d_\pm$ of the reduced CM $\text{V}^{jk}$ of the modes of interest defined as \cite{weedbrook2012}:
 \begin{eqnarray}
 d^{jk}_-&=&2^{-1/2}\sqrt{\Delta-\sqrt{\Delta^2-4\cdot\text{det}(\text{V}^{jk})}}\\
 &=&2^{-1/2}\sqrt{    (V^{jk}_{11})^2+(V^{jk}_{33})^2-2(V^{jk}_{13})^2\pm\sqrt{( (V^{jk}_{11})^2- (V^{jk}_{33})^2)^2-4(V^{jk}_{31})^2  (  V^{jk}_{11}-V^{jk}_{33})^2   }   } \label{sympletic},
 \end{eqnarray}
and the coherent information is given as \cite{Olivares2012a,Barzanjeh2015}:
 \begin{equation}
I(j\rangle k)=h(V^{jk}_{11})-h(d^{jk}_+)-h(d^{jk}_-)
\end{equation}
where:
\begin{equation}
h(x)=(x_m+0.5)\log_2(x_m+0.5)-(x_m-0.5)\log_2(x_m-0.5)
\end{equation}
with $x_m=(\tilde{d}_-^2+1/4)/(2\tilde{d}_-)$. The values for the normalized coherent information between the two optical modes $I(-\rangle +)/n^\text{out}_+$ as a function of the multiphoton cooperativities are shown in the main text.

\subsubsection{PPT and Logarithmic Negativity}
Another quantization for entanglement is based on the positivity of the partially transposed state (PPT) or Peres-Horodecki criterion which establishes that a two mode Gaussian state is entangled when the smallest symplectic eigenvalue of the partially transposed state $\tilde{\rho}_\text{EO}=(1\bigoplus T)\rho_\text{EO}$ follows 
 \begin{eqnarray}
 \tilde{d}^{jk}_-&=&2^{-1/2}\sqrt{\tilde{\Delta}-\sqrt{\tilde{\Delta}^2-4\cdot\text{det}(\text{V}^{jk})}}\nonumber\\
  &=&2^{-1/2}\sqrt{    (V^{jk}_{11})^2+(V^{jk}_{33})^2+2(V^{jk}_{13})^2-\sqrt{((V^{jk}_{11})^2-(V^{jk}_{33})^2)^2+4(V^{jk}_{31})^2  (  V^{jk}_{11}+V^{jk}_{33})^2   }   } \label{Logneg},
 \end{eqnarray}
where $\tilde{\Delta}=V^2_{11}+V^2_{33}+2V^2_{13}$ holds for the EO-CM. Therefore, the quantity  $\tilde{d}_-$ characterizes the Gaussian entanglement for any two-modes Gaussian states, where  $\tilde{d}_-<0.5$ is a condition for entanglement. Furthermore, the logarithmic negativity \cite{vidal}  uses this value to quantity the upper bound of the number of distillable entanglement of the quantum state and it is defined as \cite{logneg}:
 \begin{equation}
E^{jk}_\mathcal{N}=\text{max}[0,-\log_2(2\tilde{d}^{jk}_-)]. 
\end{equation}
In the main text the normalized logarithmic negativity $E^{+-}_\mathcal{N}/n^\text{out}_+$ for the optical outputs is shown as a function of the multiphoton cooperativity $C_1$ and $C_2$.

\subsubsection{Quantum Discord}

The correlations between two output fields of the electro-optics system can be separated between the a classical part and a quantum part, which is given by the quantum discord.
This can be calculated in terms of the  the symplectic eigenvalues of the reduced covariance matrix as given in Ref.~\cite{Barzanjeh2015}:
 \begin{equation}
D(j|k)=h(V^{jk}_{33})-h(d^{jk}_-)-h(d^{jk}_+)+h\left(V^{jk}_{11}+\frac{(V^{jk}_{13})^2(1-V^{jk}_{33})}{(V^{jk}_{33} )^2-1}\right)
\end{equation}
in the main text I show the normalized quantum discord per emitted anti-Stokes photon $D(-\rangle+)/n^\text{out}_+$  is presented as a function of the multiphotons cooperativity $C_1$ and $C_2$.

\section{$2N$-Optical Modes}
\subsection{Hamiltonian and Equation of Motion}
In a resonator described in the main text, we can find a system where many optical modes are supported and they can coupled to each other through a microwave mode.
Assuming that the spectral distance between the $N$-pump modes and their respective Stokes or anti-Stokes modes are fixed matching the single microwave resonance frequency and a non zero coupling electro-optic coupling $g_n\ne0$, the Hamiltonian in the interaction picture can be written as follows:
\begin{equation}
\hat{H}^N_\text{int}=\hbar\sum\limits_j (\hat{a}_{pj-}g_{j-}\hat{a}_{j-}^\dagger\hat{a}_\Omega^\dagger+  \hat{a}_{pj+}g_{j+}\hat{a}_{j+}^\dagger\hat{a}_\Omega)+\text{H.c.} \label{4modeha}
\end{equation}
Assuming undepleted optical pumps, the evolution of the operators in time is given as:
\begin{subequations}
\begin{align}
\dot{\hat{a}}_{j+}&=\frac{i}{\hbar}[\hat{H}^N_\text{int},\hat{a}_{j+}]=-iG_{j+}\hat{a}_\Omega\\
\dot{\hat{a}}_{j-}&=\frac{i}{\hbar}[\hat{H}^N_\text{int},\hat{a}_{j+}]=-iG_{j-}\hat{a}^\dagger_\Omega\\
\dot{\hat{a}}_{\Omega}&=\frac{i}{\hbar}[\hat{H}^N_\text{int},\hat{a}_{j+}]=-\sum\limits_NiG_{j+}\hat{a}_{j-}-\sum\limits_NiG^*_{j-}\hat{a}^\dagger_{j-}\\
\end{align}
\end{subequations}
with parametrically enhanced electro-optic couplings $\hat{a}_{pj\pm}g_{j\pm}\rightarrow G_{j\pm}$.

\subsection{Fourier Analysis}
The analysis of $2N$ optical modes entangled through a common microwave mode is an extension to the previous section. Following the the Heisenberg equation of motion for each field  and adding a given electro-optic coupling $G_n$ and loss terms to each mode $\kappa_n$, the steady state solutions for the fields in the Fourier space are given as:
\begin{subequations}
\begin{align}
\hat{a}_{j+}(\omega)&=\frac{-iG_{j+}\hat{a}_{\Omega}(\omega) +\hat{F}_{j+}(\omega)}{\Gamma_{j+}(\omega)}, \\
\hat{a}_{j-}(\omega)&=\frac{-iG_{j-}\hat{a}^\dagger_{\Omega}(\omega) +\hat{F}_{j+}(\omega)}{\Gamma_{j-}(\omega)}, \\
\hat{a}_\Omega(\omega)&=\frac{-i\sum\limits_N\left(G_{j+}\hat{a}_{j'+}(\omega)+G^*_{j-}\hat{a}^\dagger_{j'-}(\omega)\right) +\hat{F}_\Omega(\omega)}{\Gamma_\Omega(\omega)},     \label{matrixchanges}
\end{align}
\end{subequations}
where $\Gamma_{j+}(\omega)=\kappa_{j+}/2+i\omega$ with $\omega$ being the detuning from their respective mode resonance. For these equations, it was assume that fixed the expectral distance between the pump and anti- or Stoke mode is fixed at the microwave resonance frequency.  The output fields, following the side coupling relation  $\hat{a}_n^\text{out}=-\hat{a}^\text{in}_n+\sqrt{\kappa_n^e}\hat{a}^e_n$, can be written in terms of the input fields in the following way:
\begin{subequations}
\begin{align}
\hat{a}^\text{out}_{j+}(\omega)&=R_{11}(\omega)\hat{a}^e_{j+}(\omega)+R_{12}(\omega)\hat{a}^i_{j+}(\omega) +R_{13}(\omega)\sum\limits_{j\ne j'}\hat{a}^e_{j'+}(\omega)+R_{14}(\omega)\sum\limits_{j\ne j'}\hat{a}^i_{j'+}(\omega) \nonumber\\
& +R_{15}(\omega)\sum\limits_{j'}\hat{a}^{e\dagger}_{j'-}(\omega)+R_{16}(\omega)\sum\limits_{j'}\hat{a}^{i\dagger}_{j'-}(\omega)+R_{17}(\omega)\hat{a}^e_\Omega(\omega)+R_{18}(\omega)\hat{a}^i_\Omega(\omega) \\
\hat{a}^\text{out}_{j-}(\omega)&=R_{21}(\omega)\hat{a}^e_{j-}(\omega)+R_{22}(\omega)\hat{a}^i_{j-}(\omega) +R_{23}(\omega)\sum\limits_{j}\hat{a}^e_{j'+}(\omega)+R_{24}(\omega)\sum\limits_{j'}a^i_{j'+}(\omega)\nonumber \\
& +R_{25}(\omega)\sum\limits_{j'\ne j}\hat{a}^{e\dagger}_{j'-}(\omega)+R_{26}(\omega)\sum\limits_{j'\ne j}\hat{a}^{i\dagger}_{j'-}(\omega)+R_{17}(\omega)\hat{a}^e_\Omega(\omega)+R_{18}(\omega)\hat{a}^i_\Omega(\omega) \\
\hat{a}^\text{out}_\Omega(\omega)&=R_{33}(\omega)\sum\limits_{j}\hat{a}^e_{j'+}(\omega)+R_{34}(\omega)\sum\limits_{j'}\hat{a}^i_{j'+}(\omega) +R_{35}(\omega)\sum\limits_{j'}\hat{a}^{e\dagger}_{j'-}(\omega)+R_{36}(\omega)\sum\limits_{j'}\hat{a}^{i\dagger}_{j'-}(\omega)\nonumber\\
& +R_{37}(\omega)\hat{a}^e_\Omega(\omega)+R_{38}(\omega)\hat{a}^i_\Omega(\omega) \label{matrixchanges}
\end{align}
\end{subequations}
with the values for $\textbf{R}$ explcitly given as:
\begin{subequations}
\begin{align}
R_{11}(\omega)&=-1-\frac{G^2\kappa_{e,o}}{\Gamma_o^2(\omega)\Gamma_\Omega(\omega)}+\frac{\kappa_{e,o}}{\Gamma_o(\omega)}\\
R_{12}(\omega)&=\frac{G^2\sqrt{\kappa_{e,o}\kappa_{i,o}}}{\Gamma_o^2(\omega)\Gamma_\Omega(\omega)}+\frac{\sqrt{\kappa_{e,o}\kappa_{i,o}}}{\Gamma_o(\omega)}\\
R_{13}(\omega)&=-\frac{G^2\kappa_{e,o}}{\Gamma^2_o(\omega)\Gamma_\Omega(\omega)}\\
R_{14}(\omega)&=-\frac{G^2\sqrt{\kappa_{e,o}\kappa_{i,o}}}{\Gamma^2_o(\omega)\Gamma_\Omega(\omega)}\\
R_{15}(\omega)&=-\frac{G^2\kappa_{e,o}}{\Gamma^2_o(\omega)\Gamma_\Omega(\omega)}\\
R_{16}(\omega)&=-\frac{G^2\sqrt{\kappa_{e,o}\kappa_{i,o}}}{\Gamma^2_o(\omega)\Gamma_\Omega(\omega)}\\
R_{17}(\omega)&=-\frac{iG\sqrt{\kappa_{e,o}\kappa_{e,\Omega}}}{\Gamma_o(\omega)\Gamma_\Omega(\omega)}\\
R_{18}(\omega)&=-\frac{iG\sqrt{\kappa_{e,o}\kappa_{i,\Omega}}}{\Gamma_o(\omega)\Gamma_\Omega(\omega)}
\end{align}
\end{subequations}
\begin{subequations}
\begin{align}
R_{21}(\omega)&=1+\frac{G^2\kappa_{e,o}}{|\Gamma_o(\omega)|^2\Gamma^*_\Omega(\omega)}+\frac{\kappa_{e,o}}{\Gamma_o(\omega)}\\
R_{22}(\omega)&=\frac{G^2\sqrt{\kappa_{e,o}\kappa_{i,o}}}{|\Gamma_o(\omega)|^2\Gamma^*_\Omega(\omega)}+\frac{\sqrt{\kappa_{e,o}\kappa_{i,o}}}{\Gamma_o(\omega)}\\
R_{23}(\omega)&=\frac{G^2\kappa_{e,o}}{|\Gamma_o(\omega)|^2\Gamma^*_\Omega(\omega)}\\
R_{24}(\omega)&=\frac{G^2\sqrt{\kappa_{e,o}\kappa_{i,o}}}{|\Gamma_o(\omega)|^2\Gamma^*_\Omega(\omega)}\\
R_{25}(\omega)&=\frac{G^2\kappa_{e,o}}{|\Gamma_o(\omega)|^2\Gamma^*_\Omega(\omega)}\\
R_{26}(\omega)&=\frac{G^2\sqrt{\kappa_{e,o}\kappa_{i,o}}}{|\Gamma_o(\omega)|^2\Gamma^*_\Omega(\omega)}\\
R_{27}(\omega)&=-\frac{iG\sqrt{\kappa_{e,o}\kappa_{e,\Omega}}}{\Gamma_o(\omega)\Gamma^*_\Omega(\omega)}\\
R_{28}(\omega)&=-\frac{iG\sqrt{\kappa_{e,o}\kappa_{i,\Omega}}}{\Gamma_o(\omega)\Gamma^*_\Omega(\omega)}
\end{align}
\end{subequations}
\begin{subequations}
\begin{align}
R_{33}&=1-\frac{\kappa_{e,\Omega}}{\Gamma_\Omega(\omega)}\\
R_{34}&=\frac{\sqrt{\kappa_{e,\Omega}\kappa_{i,\Omega}}}{\Gamma_\Omega(\omega)}\\
R_{35}&=-\frac{iG\sqrt{\kappa_{e,\Omega}\kappa_{e,o}}}{\Gamma_\Omega(\omega)\Gamma_o(\omega)}\\
R_{36}&=-\frac{iG\sqrt{\kappa_{e,\Omega}\kappa_{i,o}}}{\Gamma_\Omega(\omega)\Gamma_o(\omega)}\\
R_{37}&=-\frac{iG\sqrt{\kappa_{e,\Omega}\kappa_{i,o}}}{\Gamma_\Omega(\omega)\Gamma_o(\omega)}\\
R_{38}&=-\frac{iG\sqrt{\kappa_{e,\Omega}\kappa_{i,o}}}{\Gamma_\Omega(\omega)\Gamma_o(\omega)}
\end{align}
\end{subequations}
For the symmetric case of the Stokes and anti-Stokes modes, we make the approximations  $\Gamma_{j\pm}\rightarrow \Gamma$, we have the output fields on resonance $\omega=0$ given by: 
\begin{subequations}
\begin{align}
\hat{a}^\text{out}_{j+}(0)&=(2\eta_o-1-2\eta_oC)\hat{a}^e_{j+}+\sqrt{4(\eta_o-\eta_o^2)}(1-C)\hat{a}_{j+}^i -\frac{2C\sqrt{\kappa_{e,o}}}{\kappa_o}\sum\limits_{j\ne n'}(\sqrt{\kappa_{e,o}}\hat{a}^e_{j'+}+\sqrt{\kappa_{i,o}}\hat{a}^i_{j'+})  \nonumber\\
&  -\frac{2C\sqrt{\kappa_{e,o}}}{\kappa_o}\sum\limits_{ j'}(\sqrt{\kappa_{e,o}}\hat{a}^{e\dagger}_{j'-}   +\sqrt{\kappa_{e,o}}\hat{a}^{i\dagger}_{j'-}  )-\frac{4iG\sqrt{\kappa_{e,o}}}{\kappa_o\kappa_\Omega}(\sqrt{\kappa_{e,\Omega}}a^{e}_{\Omega}   +\sqrt{\kappa_{e,\Omega}}a^{i}_{\Omega}  ),\\
\hat{a}^\text{out}_{j-}(0)&=(2\eta_o-1+2\eta_oC)\hat{a}^e_{j-}+\sqrt{4(\eta_o-\eta_o^2)}(1+C)\hat{a}_{j-}^i +\frac{2C\sqrt{\kappa_{e,o}}}{\kappa_o}\sum\limits_{j'}(\sqrt{\kappa_{e,o}}\hat{a}^{e\dagger}_{j'+}+\sqrt{\kappa_{i,o}}\hat{a}^{i\dagger}_{j'+}) \nonumber\\
& +\frac{2C\sqrt{\kappa_{e,o}}}{\kappa_o}\sum\limits_{j\ne j'}(\sqrt{\kappa_{e,o}}a^{e}_{j'-}   +\sqrt{\kappa_{e,o}}a^{i}_{j'-} ) -\frac{4iG\sqrt{\kappa_{e,o}}}{\kappa_o\kappa_\Omega}(\sqrt{\kappa_{e,\Omega}}\hat{a}^{e\dagger}_{\Omega}   +\sqrt{\kappa_{e,\Omega}}\hat{a}^{i\dagger}_{\Omega}  ),\\
\hat{a}^\text{out}_\Omega(0)&=(2\eta_\Omega-1)\hat{a}^e_\Omega+\sqrt{4(\eta_\Omega-\eta_\Omega^2)} \hat{a}_\Omega^i -\frac{4iG\sqrt{\kappa_{e,\Omega}}}{\kappa_o\kappa_\Omega}\sum\limits_{j'}(\sqrt{\kappa_{e,o}}\hat{a}^e_{j'+}+\sqrt{\kappa_{i,o}}\hat{a}^i_{j'+}+\sqrt{\kappa_{e,o}}\hat{a}^{e\dagger}_{j'-}   +\sqrt{\kappa_{e,o}}\hat{a}^{i\dagger}_{j'-}  ). \label{matrixchanges}
\end{align}
\end{subequations}
From these equations, the quadrature operators are calculated as defined by the Eq.~(\ref{quad}). Subsequently the square $2N+1$ covariance matrix for the entangled multiplexed frequency comb, shown in the main text, is calculated using the relation Eq.~(\ref{defcov}).

\subsection{Quantum Protocols}
\subsubsection{Teleportation}

The fidelity in a quantum channel quantifies how similiar the initial and the final modes are, and this is defined as~\cite{braunsteinkimble}:
\begin{equation}
F= \pi\int W_\text{in}(\beta)W_\text{out}(\beta)\text{d}^2\beta,
\end{equation}
where $W_\text{in}$ and $W_\text{out}$ are the initial and final Wigner functions of the unknown quantum state, respectively. Assuming the standard Braunstein-Kimble set-up \cite{braunsteinkimble} with ideal detectors for the  Bell measurements and lossless classical information transfer, the state transfer fidelity for an coherent squeezed state  $|\psi_\text{in}\rangle=|\alpha,r\rangle$ is given by~\cite{fiurasek2002,rueda2019electrooptic}:
\begin{equation}
F^\text{G}_\text{TE}(\alpha,r,C,\eta_\text{input},\eta_\text{output})=\left(4\Delta q_-^4+4\Delta q_-^2\cosh(2r)+ 1  \right)^{-1/2}\label{fidelitytele}
\end{equation}
with $\Delta q_-$ explicitly given in Eq.~(\ref{combsqueezing1})-(\ref{combsqueezing2}) as function of $\eta_k$ and $C$ for $\bar{n}$=0. The cross-quadrature squeezing with noise can be calculated from the covariance matrix using the formula:
\begin{equation}
\Delta q^{lk}_{\mp}=\sqrt{\frac{ V_{ll}V_{kk}-V^2_{lk} }{V_{kk(ll)}\cos^2(\Theta)+V_{ll(kk)}\sin^2(\Theta)  \pm V_{lk} \sin(2\Theta)]}},\label{rsq}
\end{equation}for a coveriance matrix:
\begin{equation}
\textbf{V} = \begin{bmatrix}
V_{ll}\textbf{I}&V_{lk}\textbf{Z}\\
V_{lk}\textbf{Z} &V_{kk}\textbf{I}
\end{bmatrix},
\end{equation}
with  $k\ne l$ and the cross-quadratures'  angles following: $\tan(2\Theta)=\pm 2V_{lk}/(V_{ll}-V_{kk})$ \cite{rueda2019electrooptic}. Cat states are non Gaussian states and they are given as a quantum superposition of two coherent states in the form $|\psi_\text{cat}\rangle=N(|\alpha\rangle+e^{i\phi}|-\alpha\rangle)$ with  $N=\sqrt{2+2\exp(-2\alpha)\cos(\phi)}$.
The state transfer fidelity between the nodes of the network is given by \cite{braunsteinkimble,rueda2019electrooptic}:
\begin{equation}
F^\text{cat}_\text{TE}=\frac{1}{1+2\Delta q^2_-}-\frac{1+e^{-4|\alpha|^2}-e^{-4\frac{|\alpha|^2}{1+2\Delta q^2_-}}  -e^{-8\frac{\Delta q^2_-|\alpha|^2}{1+2\Delta q^2_-}}       }{(2+4\Delta q^2_-)(1+e^{-2|\alpha|^2}\cos(\phi))^2}.
\end{equation}
The values for $\bar{n}_\Omega=0$ and $N=1,4$ are shown in the main text.
 
\subsubsection{Dense Coding}
From Ref.~\cite{Ralph2002}, I find the relations for the capacity of a coherent state channel with single quadrature encoding and homodyne detection
\begin{equation}
C_c=\log_2(\sqrt{1+4\bar{n}}),
\end{equation}
dual quadrature encoding with heterodyne detection 
\begin{equation}
C_{ch}=\log_2(1+\bar{n}),
\end{equation}
and the dense coding capacity
\begin{equation}
C_{dc}=\log_2\left(1+\frac{\eta(4\bar{n}-V_{ne}-1/V_{ne}-b+2  )}{4(\eta V_{ne}+1-\eta)}\right).
\end{equation}
as function of the mean photon number of the mode  $\bar{n}$, the detection efficiency $\eta$, and the modes cross-quadratures $V_{ne}=2\Delta q_-$ and $1/V_{ne}+b=2\Delta q_+$.  The numerical results for these function are given in terms of the multiphoton cooperativity  for $N=1$ and $N=4$  in Fig.~\ref{squeezing} in the main text.
\end{document}